\pdfoutput=1
\documentclass[10pt, twoside]{mathdoc}
\usepackage{distoperators}

\usepackage{booktabs}
\usepackage[title]{appendix}
\usepackage[shortlabels, inline]{enumitem}
\usepackage{IEEEtrantools}
\usepackage{xfrac}
\usepackage{bm,mathrsfs,dsfont}
\usepackage{graphicx}
\usepackage{caption}
\usepackage[labelformat=simple]{subcaption}
\usepackage{tikz}

\let\proglang\textsf
\let\pkg\textbf

\newcommand*{\model}{\ensuremath{\mathrm{M}}}

\newcommand*{\elpdHatCorrected}{{\ensuremath{\widehat{\mathrm{elpd}}_\mathrm{\scriptscriptstyle corrected}}}}

\newcommand{\prob}[1]{\mathbb{P}\left(#1\right)}

\newcommand*{\bias}[2]{{\mathrm{bias}\bigr(#1 \mid #2\bigl)}}

\title{Efficient estimation and correction of selection-induced bias with order statistics}
\author{Yann McLatchie}
\author{Aki Vehtari}
\affil{Department of Computer Science, Aalto University, Finland}
\makeatletter
\def\runauthor{McLatchie and Vehtari}
\def\runtitle{Efficient selection-induced bias estimation}
\makeatother

\begin{document}
\maketitle
\thispagestyle{empty}
\begin{abstract}
    Model selection aims to identify a sufficiently well performing model that is possibly simpler than the most complex model among a pool of candidates. However, the decision-making process itself can inadvertently introduce non-negligible bias when the cross-validation estimates of predictive performance are marred by excessive noise. In finite data regimes, cross-validated estimates can encourage the statistician to select one model over another when it is not actually better for future data. While this bias remains negligible in the case of few models, when the pool of candidates grows, and model selection decisions are compounded (as in step-wise selection), the expected magnitude of selection-induced bias is likely to grow too. This paper introduces an efficient approach to estimate and correct selection-induced bias based on order statistics. Numerical experiments demonstrate the reliability of our approach in estimating both selection-induced bias and over-fitting along compounded model selection decisions, with specific application to forward search. This work represents a light-weight alternative to more computationally expensive approaches to correcting selection-induced bias, such as nested cross-validation and the bootstrap. Our approach rests on several theoretic assumptions, and we provide a diagnostic to help understand when these may not be valid and when to fall back on safer, albeit more computationally expensive approaches. The accompanying code facilitates its practical implementation and fosters further exploration in this area.
\end{abstract}
\section{Introduction}
Since \citet{lindley_choice_1968} there has been a long history of Bayesian model selection using predictive performance comparison \citep[see the extensive review by][]{vehtari_survey_2012}.
In this regime, we are interested in identifying a model from a collection of models whose out-of-sample predictive performance is best, either in the hope of generalising to future unseen data \citep{vehtari_survey_2012}, or as a proxy for parameter recoverability \citep{scholz_prediction_2022}. When the number of predictors to consider is large, we might also be interested in achieving a smaller model that is capable of replicating the predictive behaviour of a larger model and use this instead for its improved interpretability, or to decrease data collection cost \citep{piironen_projective_2020}. In finite data regimes, however, there can be significant uncertainty in differentiating between models of similar performance.

Following \citet{piironen_comparison_2017}, consider a hypothetical utility for a model $\Mk$ estimated on data $D$ which we denote $u(\Mk, D)$. 
This estimate can be decomposed into two components: the true generalisation utility of the model, and the error associated with the estimate.
The authors note that for safe model selection, we do not require that the utility be unbiased.
If these utility estimates have high variance, however, then for a single data realisation, choosing the maximum utility model may lead to choosing a sub-optimal model in terms of generalisation utility.
When we select a model which performs non-negligibly worse than the oracle asymptotically, we say that we have \textit{over-fit in selection}. 
On top of this, the utility estimate for the selected model, say $\hat k$, is typically optimistic: $u(\model_{\hat k}, D) \gg \E[D]{u(\model_{\hat k})}$, where we use $\E[D]{u(\model_{\hat k})}$ to denote the expectation over datasets.
It is this over-optimism we refer to as \textit{selection-induced bias} \citep[already discussed by][]{stone_cross-validatory_1974}. 

While in general this over-optimism appears across a range of utility estimates and criteria \citep[including common information criteria;][]{burnham_model_2002}, we focus our discussion on Bayesian cross-validation \citep{Geisser:1975,geisser_eddy_predictive_1979}.

The kernel of this work is an efficient and actionable selection-induced bias estimation based on order statistics. Concretely, we:
\begin{enumerate}[nosep]
    \item discuss over-optimism when making model selection decisions based on noisy cross-validated predictive metrics;
    \item propose a lightweight selection-induced bias estimation and correction tool based on order statistics;
    \item provide a diagnostic for our estimate to understand when it is liable to be unsafe, and more computationally expensive approaches may be warranted;
    \item show empirically that correcting selection-induced bias can help expose non-negligible over-fitting; and,
    \item apply our bias correction to real and simulated data examples where we are interested in choosing between multiple candidate models, and in forward search.
\end{enumerate}
\subsection{Relation to previous work}
Predictive approaches to Bayesian model comparison have been qualitatively and quantitatively reviewed by \citet{vehtari_survey_2012} and \citet{piironen_comparison_2017} respectively. One popular method is choosing the maximum \textit{a posteriori} (MAP) model, which reduces to choosing the maximum marginal likelihood model in the case of a uniform prior over the model space, and can be performed with Bayes factors \citep{kass_bayes_1995,jeffreys_theory_1998}. The computational instability of Bayes factors, and the difficulty they have sampling from the combinatorially exploding model space, however, is well-documented \citep{han_markov_2001,oelrich_when_2020}. 
It has also been documented in literature that under the assumption of orthogonal predictors, the most predictive model usually lies closer to the median probability model \citep{barbieri_optimal_2004}.
In the context of variable selection, one might otherwise consider investigating the probability of inclusion of an individual predictor \citep{brown_multivariate_1998,narisetty_bayesian_2014}. Probability of inclusion approaches, including the median probability approach, have problems in case of collinearity as inclusion probabilities are diluted across the collinear predictors. Most contemporary work in this field has centred on defining intelligent priors over model space and model parameters \citep{ohara_review_2009}. However, \citet{piironen_projective_2020} show how such approaches provide limited understanding of the predictive improvement expected by adding a predictor, and can fail when predictors are collinear. It has further been argued that priors and model selection exist as separate components to a Bayesian workflow \citep{vehtari_survey_2012,key_bayesian_1999,bernardo_bayesian_1994}: a view we hold ourselves.

A fully decision theoretical Bayesian model selection procedure was first outlined by \citet{lindley_choice_1968}, wherein the statistician is implored to build the largest, most complex model possible which encompasses all uncertainty relating to the modelling task, and then to identify smaller models capable of closely replicating its behaviour. \citet{leamer_information_1979} motivated the use of the Kullback--Leibler (KL) divergence to quantify the distance between models' predictive distributions. \citet{goutis_model_1998} and \citet{dupuis_variable_2003} went further, and proposed that one should directly \textit{project} the reference model onto parameter sub-spaces so as to minimise the distance between the induced projected posterior predictive distribution and the posterior predictive distribution of the reference model: essentially fitting sub-models to the fit of the reference model. \citet{mclatchie_advances_2024} present practical recommendations for projection predictive model selection, and a more complete overview of recent advances therein. We are interested in cases where either the number of models is too small to justify the added complexity of projection predictive inference, or when we don't know how to perform the projection efficiently (e.g. in Student-$t$ observational family models, and some spatial models).

As a computationally efficient, intuitive, and non-parametric alternative, and one of the most common in non-Bayesian literature, statisticians compare models based on their out-of-sample performance with either cross-validation \citep[CV;][]{geisser_eddy_predictive_1979,shao_linear_1993,vehtari_bayesian_2002}, or various information criteria  \citep{burnham_model_2002,spiegelhalter_bayesian_2002,gelfand_model_1998,marriott_bayesian_2001,laud_predictive_1995,watanabe_widely_2013}. Such procedures are able to produce an almost unbiased estimate of a model's predictive ability \citep{watanabe_asymptotic_2010}, and have been reviewed extensively by \citet{arlot_survey_2010}, and by \citet{gelman_understanding_2014} in a Bayesian context. 
When using noisy predictive performance estimates such as cross-validation to select a best model, we may be over-optimistic in the utility estimate of the eventually selected model. This  selection-induced bias has been discussed in depth by \citet{ambroise_selection_2002}, \citet{reunanen_overfitting_2003}, and \citet{cawley_over-fitting_2010}.  \citet{cawley_over-fitting_2010}, in addition to demonstrating the potential risks of selection-induced bias, emphasised that this bias does not always lead to non-negligible over-fitting. In many cases, when one model is much better than others, or if there is sufficient regularisation (via priors in Bayesian context), the selection-induced bias is negligible even if the risk of selecting the ``wrong'' model is not small. We do not look to eliminate bias, but identify when it is non-negligible. In particular, we agree with past authors that selection-induced bias is not always a problem, and rather than a zero-one assessment of its existence, we endeavour to talk about negligible and non-negligible bias. 

Very little work has dealt with bias correction in model selection, and those procedures that have been previously presented require intense computation. For instance, \citet{stone_cross-validatory_1974} proposed a procedure of nested CV (which they called the ``double cross'') which is insensitive to sampling and partitioning schemes. This procedure operates with complexity $\mathcal{O}(K^2)$ in terms of the number of folds $K$: often fiendishly expensive.\footnote{For the rest of the paper, we will use $K$ to denote the number of models, but here it is used to denote the number of cross-validation folds. We will focus on the case of leave-one-out cross-validation where we have the same number of folds as data observations: $K = n$.} An alternative was provided by \citet{tibshirani_bias_2009} who made use of the bootstrap \citep{efron_introduction_1993} to correct CV estimates. While less expensive, the complexity nonetheless scales in $\mathcal{O}(K B)$ with $B$ denoting the number of bootstrap samples. Our contribution is a lightweight alternative to these procedures, requiring only the sufficient statistics from the standard CV process and \emph{no} additional computation. In other words, it has linear complexity $\mathcal{O}(K)$. 
While one run of cross-validation forward search may require several hours of computation time for the experiments seen later in this paper, and thus nested cross-validation quickly becoming infeasibly expensive, the use of order statistics requires only a matter of seconds or minutes. The bootstrap is an improvement on nested cross-validation, but as previously mentioned will not scale as well as order statistics.

While our proposal is simple, and thus cannot be expected to produce perfect results, it can serve as a useful demonstration of the dynamics of selection-induced bias: when one model is sufficiently better than other candidates then we can proceed more comfortably, and when models are close we must take care and perform a correction. 

We provide the statistician with a diagnostic for our procedure, and should one estimate it to not be safe for their task, then one can fall-back on safer albeit more computationally expensive approaches. 
While we don't claim that CV estimates are the best way to do the model selection, it can be sometimes computationally convenient and sufficiently good; our approach helps to understand when it is sufficiently good.
\subsection{Alternatives to model selection}
While predictive performance-based model selection can operate as a convenient non-parametric method of determining whether there is a significant difference between models, simply investigating parameter posteriors \textit{can sometimes} provide sharper resolution to the problem \citep{gelman_understanding_2014, wang_difficulty_2015, gelman_posterior_1996}. 
When the model parameters are interpretable, causally consistent, nested, and there is no collinearity between predictors, then we can often investigate their posteriors directly with more success.
Since the predictive distribution integrates over the uncertainty over all parameters, it may introduce more variance in the estimator than is necessary. 
For instance, we could investigate only the posterior of the extra parameter in the larger model and decide whether or not to include this parameter in our analysis based on whether its posterior differs greatly from the prior.
These procedures, as we have already discussed, however, become more difficult given predictor collinearity.

When we are not in the nested case, we might follow the suggestions of \citet{yates_parsimonious_2021} and either:
\begin{enumerate*}[label=(\arabic*), ref=\arabic*, nosep]
    \item perform Bayesian model averaging \citep[BMA;][]{hoeting_bayesian_1999} or model stacking \citep{yao_using_2018,yao_bayesian_2021,yao_locking_2023};
    \item use the largest model thoroughly-considered priors%
    ; or,
    \item consider alternative model structures \citep{garthwaite_selection_2010}.
\end{enumerate*}

BMA is an attractive initial first consideration, and has been shown by \citet{le_model_2022} to even be better in terms of prediction than model selection under some conditions. The good performance of BMA in the predictive sense has been covered previously by \citet{hoeting_bayesian_1999} and \citet{raftery_discussion_2003}.
Bayesian model averaging weights, however, can be expensive to compute, and \cite{yao_using_2018} introduce pseudo-BMA weights as a lightweight alternative 
(we discuss the connection between pseudo-BMA weights and model selection based on elpd in Appendix~\ref{sec:k2}). 
The authors discuss that when none of the models considered is the true model, this method can be flawed, and motivate stacking of the models' posterior predictive distributions as more robust alternative. This consists of identifying model weights such that the combination of densities minimises the LOO-CV mean-squared error.
More recently, \citet{yao_locking_2023} have extended this paradigm to include more exotic model combinations.
\subsection{Structure of this paper}
We begin by discussing how model selection decisions themselves are liable to induce over-confidence in the expected performance of a selected model in Section~\ref{sec:selection-induced bias}. We demonstrate how the magnitude of selection-induced bias grows with the number of candidate models in Section~\ref{sec:many-models}, wherein we present our first contribution: a novel order statistic-based tool to determine when a model selection decision is liable to induce non-negligible bias. 
In Section~\ref{sec:safe-forward-search} we demonstrate how, while a single decision might introduce negligible bias, when we make multiple sequential decisions, and with specific reference to forward search, the bias is likely to compound. We then extend our first contribution to correct for selection-induced bias in forward search, and use it as a stopping rule. 
We manifest the reliability, computational tractability, and flexibility of our correction on both simulated and real-world data exercises, wherein we compare our stopping rule with traditional alternatives.\footnote{Code to replicate the results in this paper is freely available at \url{https://github.com/yannmclatchie/model-selection-uncertainty}, and the underlying selection-induced bias estimation has been implemented in the \pkg{loo} package in \proglang{R} \citep{vehtari_loo_2023}.}
\section{Predictive model assessment, selection, and
comparison} \label{sec:selection-induced bias}
The focus of our paper is on developing a framework for modelling the predictive difference estimates between models, and specifically identifying when models are sufficiently different to safely make a model selection decision based on sufficient statistics. 

In the case of predictor collinearity the inclusion probabilities are diluted, and it is often safer to select models based on their predictive performance.
We proceed considering the log score as the default utility best suited for evaluating the predictive density \citep{bernardo_bayesian_1994}, and compute it with cross-validation \citep{geisser_eddy_predictive_1979}.
\subsection{Predictive model assessment}
\citet{vehtari_practical_2017} define the expected log point-wise predictive density (elpd) as a measure of predictive performance for a new dataset of $n$ observations:\footnote{This measure is the same as used, e.g., by \citet{geisser_eddy_predictive_1979} and \citet{watanabe_widely_2013}.}
\begin{equation}
    \elpd{\Mk}{y} = \sum_{i=1}^n \int p_t(\tilde{y}_i) \log p_k(\tilde{y}_i\mid y)\,\mathrm{d}\tilde{y}_i,
\end{equation}
where $p_t(\tilde{y}_i)$ denotes the density of the true data-generating process for $\tilde{y}_i$, and $p_k(\tilde{y}_i\mid y) = \int p_k(\tilde{y}_i\mid\theta)p_k(\theta\mid y)\,\mathrm{d}\theta$ is the posterior predictive distribution of model $\Mk$. In practice, we do not have access to $p_t(\cdot)$ and approximate the $\elpdPlain$ with its leave-one-out cross-validated (LOO-CV) estimate \citep{geisser_eddy_predictive_1979}, 
\begin{equation}
    \mathrm{elpd}_\mathrm{\scriptscriptstyle LOO}\left(\Mk \mid y\right) \coloneqq \sum_{i=1}^n \log p_k(y_i\mid y_{-i}) = \sum_{i=1}^n \log \int p_{k}(y_i \mid \theta) p_{k}(\theta \mid y_{-i}) \,\mathrm{d}\theta. \label{eq:elpd-loo}
\end{equation}
If the data $y$ are conditionally independent then \citet{gelfand_model_1992}, and later \citet{gelfand_model_1996}, demonstrate how this quantity can be efficiently estimated with importance sampling by re-weighting posterior samples $\theta^s$, for $s = 1,\dotsc,S$ indexing samples from the posterior, with weights
\begin{equation}
  r_i^s = \frac{1}{ p_k(y_i\mid \theta^s)} \propto \frac{p_k(\theta^s \mid y_{-i})}{p_k(\theta^s \mid y)}
\end{equation}
leading to the importance sampling estimate
\begin{equation}
  \elpdHat{\Mk}{y} = \sum_{i = 1}^n \log \left( \frac{\sum_{s = 1}^S r_i^s p_k(y_i\mid \theta^s)}{\sum_{s = 1}^S r_i^s} \right).
\end{equation}
\citet{vehtari_practical_2017} further demonstrate how this estimator can be stabilised and self-diagnosed with Pareto-smoothed importance sampling \citep{vehtari_pareto_2022}. Here we focused on LOO-CV, but our approach can also be used in the context of other cross-validation approaches, for example leave-future-out \citep{burkner_approximate_2020,cooper_cross-validatory_2023} and leave-one-group-out \citep{merkle_bayesian_2019}.

As well as the LOO-CV elpd point estimate, we can estimate the uncertainty in the elpd estimate under the normal approximation with standard error (SE) given by
\begin{equation}\label{eq:elpd-loo-se}
    \widehat{\mathrm{SE}}_{\mathrm{LOO}}(\Mk \mid y) = \left(\frac{n}{n - 1} \sum_{i=1}^n\left( \widehat{\mathrm{elpd}}_{\mathrm{LOO}, i}(\Mk \mid y) - \frac{1}{n} \sum_{j=1}^n \widehat{\mathrm{elpd}}_{\mathrm{LOO}, j}(\Mk \mid y)\right)^2 \right)^{1/2}
\end{equation}
where we write $\widehat{\mathrm{elpd}}_{\mathrm{LOO}, i}(\Mk \mid y) = \log p_k(y_i\mid y_{-i})$.
\citet{sivula_uncertainty_2022} discuss under which conditions this approximation is accurate.

\citet{watanabe_asymptotic_2010} demonstrated the equivalence between cross-validation and the widely applicable information criterion (WAIC).
However, in finite data regimes the error in the WAIC can be non-negligible.
As such, we don't consider the latter any further.
\subsection{Selection-induced bias}\label{sec:bias-definition}
We can then use these predictive estimates to select one model from a pool of candidates. Even if each predictive estimate is almost unbiased for each model, however, when we use these for model selection the estimate for the selected model may incur some selection-induced bias.

Selection-induced bias emerges from selecting a model from a collection of models based on a noisy CV predictive estimate which could have been shown by the true elpd to be sub-optimal with respect to the models considered.\footnote{Most of the noisiness of these estimates is borne from finite data, and not the importance sampling scheme employed.} For the rest of this paper, we suppose that we choose the model with the highest LOO-CV elpd point estimate from a collection of $K$ candidate models, 
\begin{equation}
    k^\ast = \argmax_{k = 1,\dotsc,K}\,\elpdHat{\Mk}{y},
\end{equation}
and define selection-induced bias concretely as
\begin{equation}
    \bias{\model_1,\dots,\model_{K}}{y} = \elpdHat{\model_{k^\ast}}{y} - \elpd{\model_{k^\ast}}{y}.\label{eq:selection-induce-bias}
\end{equation}
We then wish to understand under which situations choosing the model with the best LOO-CV elpd estimate may lead to non-negligible over-optimism in expected LOO-CV elpd (selection-induced bias), and how bad this bias is as a function of model similarity and the number of candidate models.

If the bias is small compared to the SE of the normal approximation to the elpd LOO estimate (Equation~\ref{eq:elpd-loo-se}), we say that there is negligible selection-induced bias.
However, \citet{sivula_uncertainty_2022} show that the standard deviation of the normal approximation may be under-estimated when the models are very similar. 
As such, extra care should be taken when the absolute difference in elpd between models is small (smaller than $4$, say; see Appendix~\ref{sec:k2}). 
Depending on the application, there may be additional information pertaining to what is a practically relevant magnitude of the difference between model performances.

Rather than measure the performance of any model selection decision in terms of whether or not we choose the correct model (since the notion of a true model is difficult to justify in practice), we instead discuss how far we are from selecting the oracle model \citep[in terms of minimal excess loss in expectation, as defined by][we will return to this discussion in Section~\ref{sec:many-models}]{arlot_survey_2010}. In a word, we interest ourselves in the expected selection-induced bias injected by a given model selection decision, aim to minimise this as much as possible, and correct it when it is non-negligible. When we expect a model selection decision to induce only negligible amounts of selection-induced bias, we consider it a ``safe'' decision.
\subsection{When is one model clearly better than another?}\label{sec:motivate}
We begin by illustrating the $K = 2$ regime, in which we have only two candidate models. A typical question arises: is one of these models better than the other? These two models may be achieved with different priors, be nested (as in one iteration of forward search, say), or may be completely non-nested. Not only are we interested in discerning when it is safe to select one of the two models over the other, we would also like to be able to say something about the potential danger associated with simply selecting the better model in terms of LOO-CV elpd. In this section, we firstly present the probability of selecting the wrong model under some oracle knowledge, before discussing the selection-induced bias arising from this decision when the models are indistinguishable. We interest ourselves later on in this paper primarily in discerning when models are indistinguishable in terms of their predictive performance.

Consider two models, $\Ma$ and $\Mb$, and suppose we know that $\Ma$ is the oracle model. We follow the definition of \citet[Section 2.2;][]{arlot_survey_2010} in that the oracle model is the model having the minimum excess loss in expectation over multiple datasets from a collection of models. In particular, it is possible for the oracle model to differ from the model containing the truly relevant predictor set (for simplicity we call that the ``true model'') in cases where that model can over-fit. In general, we don't always need to choose the oracle model; the excess loss can just be negligible. It is often more useful to discuss this difference in terms of a continuous scale, rather than focus on the (often unrepresentative) case of either selecting the oracle model or not.

The structure of these two models does not influence model selection since we only consider their elpds. Denote the difference between the models' theoretical elpds as
\begin{equation}
    \Delta\elpd{\Ma,\Mb}{y} = \elpd{\Ma}{y} - \elpd{\Mb}{y}.
\end{equation}
As previously mentioned, we are most interested in the situation where no model is truly better than any other, that is that $\Delta\elpd{\Ma,\Mb}{y} = 0$. When this is the case, the estimated $\Delta\elpdHat{\Ma,\Mb}{y}$ represents pure selection-induced bias, regardless of the model chosen.

Consider the following (purely illustrative) setting: assume the distribution of elpd \textit{point estimates}, $\Delta\elpdHat{\Ma,\Mb}{y}$, is a Gaussian centred on the mean $\mu$ and having variance $\sigma^2 = 1$ \citep[similar to the setup employed by][]{sivula_uncertainty_2022}. From this distribution, we ``sample'' one observation of the LOO-CV elpd difference, denoted\footnote{In practice, the data induce a LOO-CV point estimate. For the purposes of this demonstration, we assume that we can sample from the distribution of LOO-CV point estimates directly. Throughout this paper it is the distribution of these point estimates that we interest ourselves in primarily.} 
\begin{equation}
    \Delta\elpdHat{\Ma,\Mb}{y}\coloneqq\Delta\widehat{\mathrm{elpd}}_{a,b} \sim p(\Delta\widehat{\mathrm{elpd}}_{a,b}) \equiv \normal(\mu, 1^2). \label{eq:delta-elpd-ab}
\end{equation}
When this observed elpd difference $\Delta\widehat{\mathrm{elpd}}_{a,b}$ is negative and we subsequently select $\Mb$ over $\Ma$ based on the LOO-CV estimate then we are at risk of inducing non-negligible selection-induced bias. Naturally, since $\Ma$ is the oracle model, the true elpd difference is non-negative.

\begin{figure}[!t]
    \centering
    \includegraphics{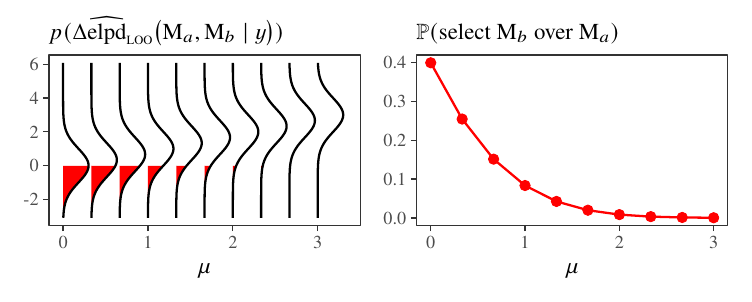}
    \caption{The left plot shows illustrative LOO-CV elpd difference distributions betwee the true model $\Ma$ and alternative model $\Mb$, as described in Equation~\ref{eq:delta-elpd-ab}. If we observe a LOO-CV estimated elpd difference the region coloured in \textcolor{red}{red}, then we are at risk of selecting the wrong model. The right plot shows this probability across different hypothetical mean values as computed in Equation~\ref{eq:gaussian-risk}.}
    \label{fig:gaussian-bias}
\end{figure}
In this toy example, we can compute the theoretical probability of selecting the sub-optimal model, with the knowledge that $\Ma$ is the oracle model, as
\begin{equation}
    \pr{\text{select }\Mb\text{ over }\Ma} = \int_{-\infty}^0 \Delta\widehat{\mathrm{elpd}}_{a,b} \cdot p(\Delta\widehat{\mathrm{elpd}}_{a,b}) \,\mathrm{d}(\Delta\widehat{\mathrm{elpd}}_{a,b}), \label{eq:gaussian-risk}
\end{equation}
where $p(\Delta\widehat{\mathrm{elpd}}_{a,b})$ is the oracle distribution of the elpd difference between the two models (omitting the dependence on the data $y$ for brevity). We visualise this theoretical illustration in Figure~\ref{fig:gaussian-bias}, where we let the mean slowly grow away from zero, keeping the standard deviation the same. The probability of selecting the wrong model then is greatest for a given $\sigma$ when the two models are most similar ($\mu$ closest to zero), and decays more slowly with an increasing mean of the elpd difference distribution.

The selection-induced bias is likely to be non-negligible when it is much larger than the variation in the LOO-CV point estimates. In terms of Figure~\ref{fig:gaussian-bias}, if the selection-induced bias is less than the illustrative $\sigma^2 = 1$, then it is less critical to model selection than if it were much larger. In a word, selection-induced bias is said to be negligible if it is dominated by the variation in the LOO-CV point estimates.

\begin{figure}[!t]
    \centering
    \includegraphics{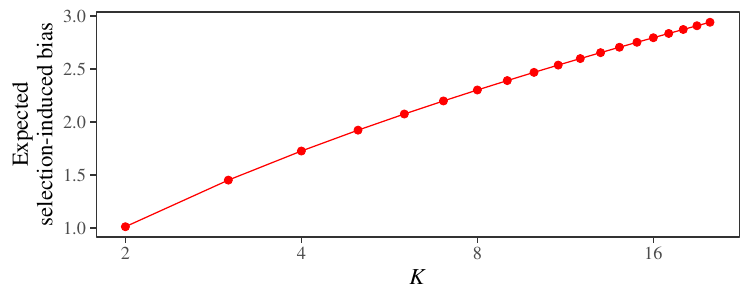}
    \caption{Theoretical illustration. The hypothetical expected selection-induced bias in the presence of $K$ candidate models. While the bias is negligible for $K=2$, it can grow rapidly with the number of candidate models. We discuss this trend later.}
    \label{fig:order-stat-theory}
\end{figure}

Nonetheless, in our experience (outwith this paper) we are able in most cases to simply choose the more complex model fitted with reasonable priors since the bias is negligible in the $K=2$ case. However, as the number of effectively equivalent models grows \citep[e.g., if the number of candidate models grows exponentially with the number of observations $n$, or if the number of predictors $p$ is much larger than $\log(n)$ as described by][]{gelman_bayesian_2014} the potential bias grows too. 
While selection-induced bias remains negligible in the two model case, it can balloon as $K$ grows. The specifics of the rate shown in Figure~\ref{fig:order-stat-theory} is described later in this paper. This many model regime is the more interesting case.
\section{Bias grows with the number of candidate models}\label{sec:many-models}
In the regime of many candidate models, when none is truly more predictive than any other, the probability of one model \textit{appearing} better than the rest grows with the number of models being compared. This is true even if independent test data would show that there is in fact no difference from it to the true model (or that it is in fact worse). As this probability grows, so too does the magnitude of selection-induced bias. In this section, we first show how these high risk regions behave as the number of models grows, and propose a selection-induced bias estimation scheme based on order statistics to identify when the bias is no longer negligible.
\subsection{Empirical demonstration}
We fit a series of nested linear models to $n = 100$ data points sampled from 
\begin{align}
    y&= X\beta + \epsilon \label{eq:many-K-experiment-1} \\
    \epsilon&\sim\normal(0,\,\sigma^2 I), \label{eq:many-K-experiment-2}
\end{align}
where $I$ is the $n\times n$ identity matrix, and $0\in\mathbb{R}^n$ is the $n$-dimensional zero vector. The data $X_i = (1,X_{i,1},\dotsc,X_{i,K - 1})$ are $K$ dimensional, with $X_{i,1},\dotsc,X_{i,K - 1}\iidsim\normal(0,\,1)$, and the true regression coefficients take the form $\beta = (1,\beta_\Delta,0,\dotsc,0)\in\mathbb{R}^{K}$. The variance of the residuals, $\sigma$, is chosen to vary to the signal-to-noise with $\beta_\Delta$ while keeping the marginal variance constant. In particular, we only consider $\beta_\Delta < 1$, and set $\sigma^2 = 1 - \beta_\Delta^2$. We will then fit a baseline model comprising only the intercept term and denoted $\model_{\mathrm{base}}:\,y_i\mid\beta_1,\tau\sim\normal(\beta_1 ,\, \tau^2)$. We compare this to the remaining $K-1$ one-predictor models of the form
\begin{equation}
    \model_{k}:\,y_i\mid\beta_1,\beta_{k},\tau\sim\normal(\beta_1 + X_{i,k}\beta_{k},\, \tau^2),
\end{equation}
for $k = 2,\dotsc,K$ so that model $\model_2$ alone contains the two predictor set, and the scale factor $\tau$ is treated as an unknown parameter across all models. Naturally, letting $\beta_\Delta = 0$ collapses all models to be equivalent.\footnote{We fit the model using independent $\normal(0,10^2)$ priors over the regression coefficients, and a half-normal prior over $\tau^2$ with mean zero and unit variance.}

\begin{figure}[!t]
    \centering
    \includegraphics{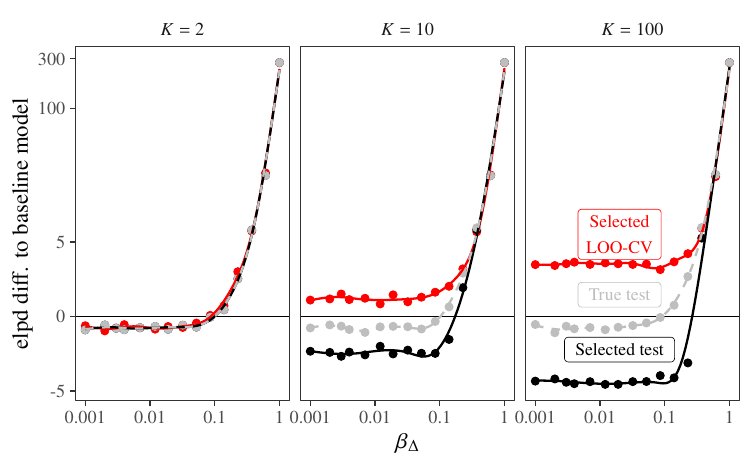}
    \caption{Simulated nested linear regression example. The \textcolor{red}{red} lines indicate the best LOO-CV elpd difference of $K$ candidates from a baseline model (the horizontal zero line), and the \textcolor{black}{black} lines are the test elpd difference. In \textcolor{gray}{grey} is the test elpd of the model fit with the true predictors. 
    The region between the \textcolor{red}{red} and black curves is selection-induced bias, whereas the region between the oracle model (this is the zero line up to $\beta_\Delta\approx 0.3$, after which it is the \textcolor{gray}{grey} line) and \textcolor{black}{black} curves is what we refer to as over-fitting.
    The magnitude of the selection-induced bias grows with $K$.}
    \label{fig:many-K}
\end{figure}

We vary the true $\beta_\Delta$ an $\epsilon$ distance from zero. For each value of $\beta_\Delta$, we investigate the LOO-CV elpd difference point estimates of the $K - 1$ models from the baseline model, and identify the model with the highest (best) cross-validated score. This is then compared to its test elpd, and to that of the model with the known true predictor set (which we call the ``true model'', even if we do not use the true values of the predictors). We show this for different numbers of candidate models $K$ and along $\beta_\Delta$ in Figure~\ref{fig:many-K}. As previously mentioned, the selection-induced bias (difference between the LOO-CV elpd and the test elpd of the selected model) is negligible when $K = 2$, but grows with $K$. For $K=100$, it would be unsafe to make a decision based on LOO-CV elpd point estimates up until the theoretical $\beta_\Delta$ grows above $\approx 0.3$, as the selection-induced bias is non-negligible. 
There is no formal threshold for non-negligible bias, but in Appendix~\ref{sec:k2} we show how the difference between elpd estimates can be connected to the notion of non-zero weights in model averaging. 
The magnitude of the bias also grows with $K$, seen in the gap between the LOO-CV elpd and test elpd of the selected model. 

When the true value of $\beta_\Delta$ is low, the model containing the known true predictors does not out-perform the baseline model in terms of test elpd. It is in these situations where the difference between the oracle model and so-called true model becomes apparent. In this case, the baseline model has the lowest excess loss in expectation over data realisations. When the true model performs worse than the oracle model, but we proceed to select the former, we have over-fitting. This excess loss from the oracle model is the difference from the baseline model to the selected model's test elpd. As $\beta_\Delta$ grows sufficiently large to consistently select the true model, this tends to zero, as too does selection-induced bias. The region between the LOO-CV and test elpd of the selected model is the selection-induced bias, while the region between the oracle elpd and the test elpd of the selected model is over-fitting. In this example, the magnitude of over-fitting is negligible across all values of $\beta_\Delta$.

Since there exist situations where the true model performs worse than the oracle model, we find some justification for our previous discussion in Section~\ref{sec:bias-definition}.
Namely, we should be more interested in identifying models which are close to the oracle model rather than retrieving the true model (since the two are not equivalent).
Indeed, finding the true predictors is not useful if it worsens prediction. 
These true predictors we ``fail'' to select are likely to have a coefficient magnitude below the detection limit, as illustrated when we increase $K$ for a small (but non-zero) $\beta_\Delta$.

Knowing then that the risk of selection-induced bias exists, and not knowing the true value of $\beta_\Delta$, we presently concern ourselves with producing a selection-induced bias estimation procedure geared towards identifying high risk regimes in the presence of many candidate models by modelling the distribution of LOO-CV elpd difference estimates.
\subsection{Our proposal: modelling the elpd differences}\label{sec:order-stat-proposal}
Suppose now that we have $K$ candidate models, none of which is better than any other. The best LOO-CV elpd difference point estimate from a baseline model may appear, purely by chance, to be greater than zero. By modelling the underlying distribution of the elpd difference point estimates as iid draws from a Gaussian distribution centred on zero (so that none is truly better than the baseline model) with some variance $\sigma_K^2$, $\Delta \elpdHatPlain^{(i)} \sim\normal(0,\sigma_K),\,i = 1,\dotsc,K$, we can leverage order statistics to produce a selection-induced bias estimate.

\citet{blom_statistical_1960} posited an approximation to the expected maximum from $K$ iid samples from a standard normal distribution, denoted $S^{(K)}$ \citep[later developed by][]{royston_algorithm_1982,harter_expected_1961}:
\begin{equation}
    S^{(K)}=\Phi^{-1}\left( \frac{K - \alpha}{K - 2\alpha + 1} \right) . \label{eq:max-approx}
\end{equation}
The formulation allows for an approximation parameter, $\alpha$.
In the limit as $n\to\infty$, the expectation of the maximum order statistic lies within the interval between levels $\alpha = 0.39$ and $\alpha = 0.5$. As such, for maximum safety we will use $\alpha = 0.5$ herein.

In the presence of $K$ candidate models, they are practically equivalent if the maximum LOO-CV elpd difference point estimate to the baseline model is less than
\begin{equation}\label{eq:order-stat-heuristic}
    S^{(K)} \hat{\sigma}_K.
\end{equation}
We estimate $\hat{\sigma}_K$ as the maximum likelihood estimator of the standard deviation of a half-normal distribution fit to the upper half-tail of elpd difference point estimates (greater than their median, $\hat{m}_K$), 
\begin{equation}
    \hat{\sigma}_K = \sqrt{\frac{2}{K}\sum_{k=1}^{K}\mathbf{1}_{\elpdHatPlain^{(k)} \geq  \hat{m}_K}\left(\Delta\elpdHatPlain^{(k)}\right)\cdot\left(\Delta\elpdHatPlain^{(k)} - \hat{m}_K\right)^2}.
\end{equation}
We denote by $\mathbf{1}_A(x)$ the indicator function over a set $A \subset X$ at $x\in X$. This is not yet an estimate of the selection-induced bias, but the highest LOO-CV elpd difference expected from a collection of $K$ models, none of them truly more predictive than any other. In particular, this is an estimate of the expected distance between the zero line and the red line in Figure~\ref{fig:many-K}. We discuss later in Section~\ref{sec:forward-order-stat} how one might use this as the basis for such an estimate of the selection-induced bias (the difference between red and black line in Figure~\ref{fig:many-K}).

We fit a half-normal distribution to the upper half of the observed elpd difference point estimates instead of a fitting a full normal distribution over all of them because (even if none of them is truly relevant) the distribution of elpd difference point estimates is likely to be slightly skewed. A similar result has previously been noted by \citet{sivula_uncertainty_2022}, who showed the \textit{full} elpd difference distribution (not just the point estimates) to be more skewed the more similar the models being compared are. We interest ourselves only in the right tail of the distribution of point estimates as it is the more relevant to the maximum order statistic. We discuss the validity of this stabilisation technique and how to diagnose it in Section~\ref{sec:diagnostics}.

\begin{figure}[!t]
    \centering
    \includegraphics{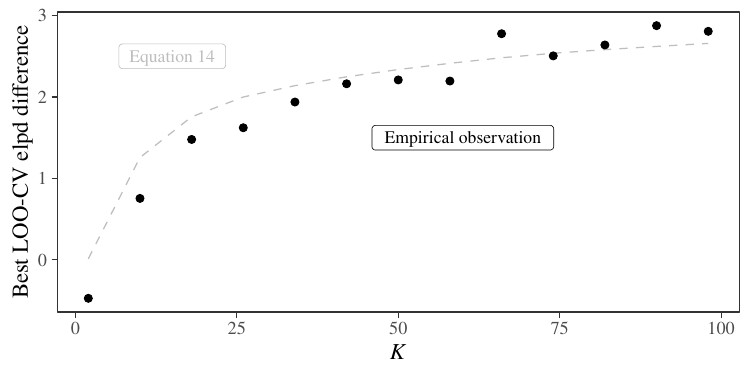}
    \caption{Simulated nested linear regression example. The mean best LOO-CV elpd difference of $K$ candidate models from the baseline model (black dots) over $100$ data realisations when $\beta_\Delta = 0$, and the expected best observed LOO-CV elpd difference of $K$ models as computed according to Equation~\ref{eq:order-stat-heuristic} (dashed \textcolor{gray}{grey} line).}
    \label{fig:order-stat}
\end{figure}

In Figure~\ref{fig:order-stat}, we fix $\beta_\Delta = 0$ and vary $K$ in the experiment design previously described by Equations~\ref{eq:many-K-experiment-1}--\ref{eq:many-K-experiment-2}. For each value of $K$, we investigate the best observed LOO-CV elpd difference to the baseline model. Since we have fixed $\beta_\Delta = 0$, we know that no model is theoretically better than any other. As such, these empirical LOO-CV elpd differences are just random noise. We repeat the experiment $100$ times for each value of $K$ and show the mean of the empirically observed best LOO-CV elpd difference in each regime compared to the expected best observed LOO-CV elpd difference as computed according to Equation~\ref{eq:order-stat-heuristic}. This theoretical boundary grows quickly for $K$ less than $10$, before slowing. Our order statistics-based equation closely matches empirical observations in this example.

A second empirical demonstration of Equation~\ref{eq:order-stat-heuristic} is provided in Appendix~\ref{sec:order-stat-demo} with reference to a real data study on the effect of a poverty reduction intervention on infant brain activity.
\subsection{Diagnosing the theoretic assumptions}\label{sec:diagnostics}
We have made a number of assumptions thus far in our analysis. In particular, we have assumed that the elpd difference distributions of models of equivalent predictive performance are:
\begin{enumerate}[nosep]
    \item normally distributed around zero (or at least half-normal above zero);  and, 
    \item independent and identically distributed.
\end{enumerate}
It is known that the empirical elpd difference distributions are slightly skewed, with heavier right tails \citep[this was seen in empirical results not shown here, and has previously been discussed by][]{sivula_uncertainty_2022}. This behaviour becomes more pronounced as models become closer together (through the use of regularising priors for example), and such correlation may appear as models make more similar predictions. It is possible then that the distribution of elpd difference point estimates (which we interest ourselves in) may also be slightly skewed. As such, the normal approximation may not be representative of the underlying distribution, and the half-normal approximation is better able to capture the characteristics of the right tail. Indeed, it is this right tail that is of most interest to us, as it contributes most to the maximum order statistics. 

It would be possible to fit a sup-Gaussian distribution to this tail with more parameters, e.g. a half-Student-$t$, but as shown empirically, the half-normal works sufficiently well. Since it has fewer parameters we expect it to be more robust, and from the half-normal we get information from the bulk of the distribution. For example we could also have fit a Pareto distribution to the tail as done by \citet{vehtari_pareto_2022}, or a Student-$t$ distribution. We would, however, then have to estimate more moments of the distribution, which we found to be unstable in our experiments. If the utility distributions are truly skewed, we expect our normal approximation to inject some bias, but if we use a more flexible approximation there would be higher variance.
In general there is no obvious choice of optimal approximation to the utility distributions, and indeed this represents an interesting direction of future investigation.

We are also able to diagnose when our half-normal approximation is likely to fail: when the right tail of the empirical distribution has infinite variance. We follow \citet{vehtari_pareto_2022} and first identify a tail cut-off point of the empirical elpd difference point estimate distribution in line with \citet{scarrott_review_2012}, and subsequently fit a generalised Pareto distribution to this tail. \citet{zhang_new_2009} present an efficient Bayesian procedure for estimating the parameters of the distribution. In particular, the $\hat{k}$ parameter of the generalised Pareto distribution is related to its number of fractional moments. We follow the rule of thumb presented by \citet[Section 3.2.4., Equation 13]{vehtari_pareto_2022} in establishing the minimum required value of $\hat{k}$ to be confident that the right tail has finite variance when comparing 10 or more models.\footnote{We require the number of models to be at least 10 since at least 5 tail samples are needed to reliably estimate the $\hat k$ parameter.}

When this Pareto $\hat k$ value is too large, it suggests that the elpd difference point estimates are likely from a distribution with infinite variance. It is at this stage that one should fall back on safer, albeit more computationally expensive approaches: the double-cross \citep{stone_cross-validatory_1974} or the bootstrap \citep{tibshirani_bias_2009}.

Elsewhere, we know that the distribution of the maximum order statistic (of which we interest ourselves only in the mean) is slightly skewed. Because of this skewed-ness, it is possible that the mean alone (which we have denoted $S^{(K)}$) under-estimates the maximum of LOO-CV elpd difference estimates.

When the model space grows too large to consider each possible model at once, and if the models can be nested, then one might favour some efficient search through the model space. Such searches come with their own risks. We presently move from the realm of single decisions towards sequential model selection.
\section{Compounding the bias through forward search}\label{sec:safe-forward-search}
While selection-induced bias might remain negligible when we need only select one model from a group of models once, when we compound decisions, the magnitude of the bias is likely to compound too. For instance, in forward search, we begin with a baseline model (usually the intercept-only model) and then consider all models containing one predictor and the intercept term. Having fit all so-called ``size-one'' candidate models, we choose the one with the highest LOO-CV elpd point estimate. We now perform another forward step, and fit all models containing the intercept, the predictor chosen previously, and a second additional predictor. Once more we select the size-two model based on the highest LOO-CV elpd point estimate. We repeat this either until either some stopping criterion is met, or we have exhausted all available predictors.

We can easily implement various other search heuristics, such as exhaustive search \citep{galatenko_highly_2015}, backward search \citep{nilsson_artificial_1998}, and stochastic search \citep{george_variable_1993,ntzoufras_stochastic_2000}. However, we proceed with forward search alone as it 
affords the possibility of an efficient stopping rule \citep{harrell_regression_2001} and intuitive compounded bias estimation (which we present in Section~\ref{sec:forward-order-stat}).

At the beginning of this search, it is likely that there exist at least some predictors that improve on the predictive performance of the baseline model, but there comes a point after which the addition of further predictors affords the statistics no theoretical predictive performance. We will refer to this as the ``point of predictive saturation'' herein. We wish to identify this point, and estimate the bias induced by continuing the search beyond it. Under appropriate prior consideration, the full model is considered the oracle model. If one witnesses a sub-model having better test performance than the full model, then we recommend they reconsider their priors. In practice, we do not need to identify the oracle model size, only a sub-model having negligible excess loss compared to the oracle model.

When we observe a separation of the LOO-CV and test elpd along the search path, we can conclude that there is non-negligible selection induced bias in the performance estimates: purely due to the predictor ordering achieved by LOO-CV based search, we are led to believe that some sub-models can outperform larger models. This is seen by the elpd estimate at the bulge being greater than that of the final full model. In practice, \citet{mclatchie_advances_2024} argue that should one observe such a bulge in the forward search path, then one should cross-validate over search paths (something akin to the double-cross), but only up to the model size inducing this bulge. This is because cross-validating over forward search paths can be expensive, and we understand that the model size of maximal selection-induced bias is likely greater than the model size of theoretic predictive saturation.

\begin{figure}[!t]
    \centering
    \includegraphics{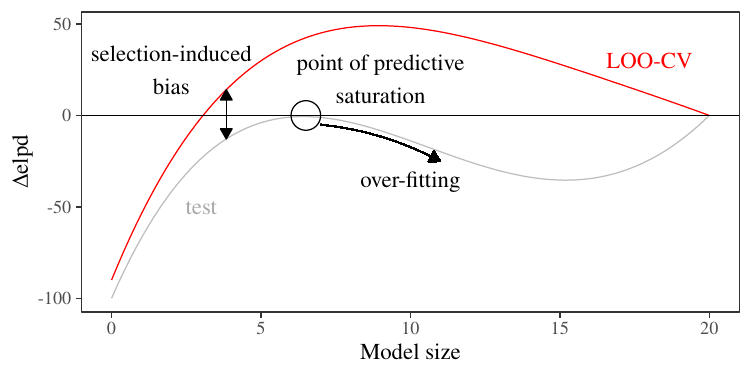}
    \caption{Illustrative sketch. 
    In \textcolor{red}{red} we show the LOO-CV elpd difference to the full (reference) model along growing model size, and in \textcolor{gray}{grey} the out-of-sample elpd difference. 
    Selection-induced bias is the divergence of the \textcolor{red}{red} line from the \textcolor{gray}{grey}.
    Over-fitting is the drop in predictive performance with increased model size following the point of predictive saturation.}
    \label{fig:forward-over-fit}
\end{figure}

The dynamics of forward search can be seen in the relationship between the predictive criterion (the elpd, say) and the model sizes along the so-called ``search path''. The plot typically shows an increasing elpd curve, indicating that the addition of further predictors improves predictive performance. We expect the curve to flatten after a given model size, indicating that the addition of further predictors does \textit{not} improve the criterion any longer: the point of predictive saturation. However, it is possible that in the finite data regime, and under model misspecification, we instead witness a bulge along this search path, such as that shown in Figure~\ref{fig:forward-over-fit}. 
The test curve in Figure~\ref{fig:forward-over-fit} reaches the point of predictive equivalence and then dips down with increased model size. This is \textit{not} necessarily selection-induced bias, but could also be over-fitting. Indeed, we aim to correct selection-induced bias to identify non-negligible over-fitting in forward search.

In order to correct selection-induced bias and estimate the location of the point of predictive saturation (and the bias induced by continuing beyond it) we introduce the following application of our proposal from Section~\ref{sec:order-stat-proposal}.

\subsection{Correcting for selection-induced bias in forward search}\label{sec:forward-order-stat}
We can estimate the selection-induced bias at model size $k$ in a forward search path, and correct the LOO-CV elpd difference estimate, denoted $\Delta\elpdHatPlain^{(k)}$, accordingly by
\begin{equation}\label{eq:corrected-elpd-diff}
    \Delta\elpdHatCorrected^{(k)} = \begin{cases} \Delta\elpdHatPlain^{(k)} - \widehat{\text{bias}}^{(k)},\,\mathrm{if }\,\vert\Delta\elpdHatPlain^{(k)}\vert < S^{(k)} \hat\sigma_k \\
    \Delta\elpdHatPlain^{(k)},\,\mathrm{otherwise},
    \end{cases}
\end{equation}
where $S^{(K)} \hat\sigma_K$ is as defined in Section~\ref{sec:order-stat-proposal}. Intuitively, if we detect non-negligible selection-induced bias, then we correct the LOO-CV elpd difference by an estimate of the selection-induced bias induced by step $k$. 

We produce this estimate of selection induced bias, denoted $\widehat{\text{bias}}^{(k)}$, building on Section~\ref{sec:order-stat-proposal} as $\widehat{\text{bias}}^{(k)} = 1.5\times S^{(k)} \hat\sigma_k$. 
The nature of this estimate stems from \citet[Table 3]{watanabe_asymptotic_2010}, who discussed how cross-validated scores and the average Bayes generalisation loss can exhibit strong negative correlation (albeit demonstrated with only one example). Theoretically this is due the removal of an (influential) observation from the training set skewing the posterior predictive distribution away from the true data-generating distribution, and its re-appearance in the test set seeming like an outlier. As such, the correction of LOO-CV elpd estimates warrants not only a correction to the median (the difference between red line and zero line in Figure~\ref{fig:many-K}), but a negatively-correlated correction (the difference between red and black lines in Figure~\ref{fig:many-K}). In a word, the corrected elpd difference value should be a reflection across the median of candidate elpd differences. The choice of magnitude of $1.5$ is reflective of the magnitude of the negative correlation of approximately $-0.8$ noted by \citet{watanabe_asymptotic_2010}. However, since this number was achieved only over one example, we understand that there can be instances where this correlation can be closer to $0$ or $-1$. 
As such, we show repetitions of some later experiments using values of $\{1, 1.5, 2\}$ when estimating the bias in Appendix~\ref{sec:alt-correlations}.

Cumulatively summing these corrected difference estimates,
\begin{equation}\label{eq:corrected-elpd}
    \elpdHatCorrected^{(k)} = \elpdHatPlain^{(0)} + \sum_{i=1}^k \Delta\elpdHatCorrected^{(i)},
\end{equation}
starting at  the base LOO-CV elpd estimate at model size $0$, denoted $\elpdHatPlain^{(0)}$, allows us to estimate the bias induced by continuing beyond the point of predictive saturation.
As such, our proposition comprises two parts: an order statistics-based estimation of selection-induced bias (from Section~\ref{sec:order-stat-proposal}), and an associated correction term.
\subsection{A simulated example}\label{sec:toy-example}
We can visualise the performance of our correction through a simple simulated data experiment: we generate data according to 
\begin{align}
    x &\sim \normal(0,R)  \\
    y &\sim \normal(w^Tx,\sigma^2), 
\end{align}
where the number of predictors is set to $p = 100$, $\normal(\mu, \Sigma)$ denotes a $p$-dimensional normal distribution with mean vector $\mu$ and covariance matrix $\Sigma$. The matrix $R \in \mathbb{R}^{p\times p}$ is block diagonal, with each block having dimension $5 \times 5$. Each predictor has mean zero and unit variance and is correlated with the other four predictors in its block with coefficient $\rho$, and uncorrelated with the predictors in all other blocks. We will investigate regimes of zero and high correlation: $\rho=\{0, 0.9\}$. Further, the weights $w$ are such that only the first $15$ predictors influence the target $y$ with weights $(w_{1:5},w_{6:10},w_{11:15}) = (\xi,0.5 \xi,0.25 \xi)$, and zero otherwise. 
We set $\xi = 0.59$ and $\sigma^2=1$ to fix $R^2 = 0.7$. We simulate $n=\{100, 200, 400\}$ data points according to this data-generating process (DGP). Across all six $n$ and $\rho$ regimes, we perform forward search based on the maximum LOO-CV elpd point estimate of the candidate model, fitting each model using independent Gaussian priors over the regression coefficients (which we understand to be liable to over-fit), and comparing them to a reference model fit with the sparsity-inducing R2D2 prior 
(which is discussed in more detail in Section~\ref{sec:priors} and Appendix~\ref{sec:r2d2}).

In Figure~\ref{fig:normal-toy-example} we visualise the performance of our 
bias correction across these six regimes in terms of the mean log predictive density (mlpd) computed on $\tilde{n}$ independent test observations, $\tilde{y}_{1:\tilde{n}}$, as
\begin{equation}
    \mathrm{mlpd}\left(\Mk\mid\tilde{y}\right) = \frac{1}{\tilde{n}}\sum_{i=1}^{\tilde{n}} \log p_k(\tilde{y}_i\mid \theta), \label{eq:mlpd}
\end{equation}
which admits a LOO-CV difference estimate similar to the elpd \citep{piironen_comparison_2017}. 
We do not correct the LOO-CV differences beyond their LOO-CV bulge since at this point it is already clear that we have over-fit the data.

We observe from these plots two positive results and one negative:
\begin{enumerate}[nosep]
    \item the corrected mlpd does not greatly surpass the mlpd of the of the reference model, contrary to the LOO-CV mlpd;
    \item the corrected mlpd estimates the point of predictive saturation close to the true point of predictive saturation on test data; however,
    \item the corrected mlpd of the search path can under-estimate the true selection-induced bias slightly in cases of severe over-fitting (most notably when $n = p$).
\end{enumerate}
This suggests that a simple order statistic model is able to represent selection-induced bias well enough to be useful. Naturally, the quality of our correction improves with data size and the number of candidate models. 

The approximation performs best in the larger data settings. It would further appear that, in this example, once $n > 4p$ the degree of selection-induced bias is negligible. 
This could be due in part to the negative correlation between Bayes cross-validation and the average Bayes generalisation loss  \citep{watanabe_asymptotic_2010}: even if we know the underlying distribution of LOO-CV elpd estimates, the asymptotic theoretical elpd distribution is likely negatively correlated with it (and worse in absolute value). Such correlation is naturally exaggerated in smaller data regimes, as the relative influence of data points grows, and our proposal under-estimates the true selection-induced bias. Despite under-estimating the bias slightly in the $n = p$ case, it appears to correct enough of the selection-induced bias to accurately and efficiently identify over-fitting across all regimes.

As previously mentioned, it is not clear how one would motivate any particular selection-induced bias correction beyond what we have proposed 
(see Section~\ref{sec:discussion} for more discussion, and Appendix~\ref{sec:alt-correlations} for a repetition of the experiments using different levels of correction), and we rather recommend that one endeavours instead to mitigate against non-negligible over-fitting in the first instance. One such way is with a more principled application of the prior. 
\subsection{Reducing selection-induced bias with priors}\label{sec:priors}
\begin{figure}[!t]
     \centering
     \begin{subfigure}{\linewidth}
         \centering
         \includegraphics{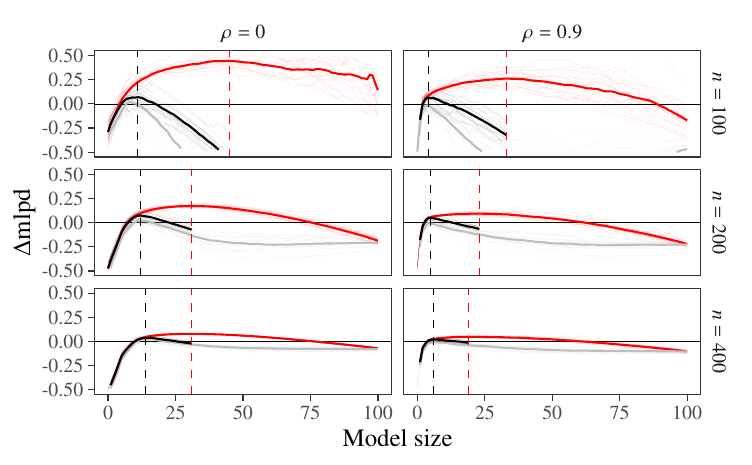}
         \caption{Using independent Gaussian priors over regression coefficients.}
         \label{fig:normal-toy-example}
     \end{subfigure}
     \hfill
     \begin{subfigure}{\linewidth}
         \centering
         \includegraphics{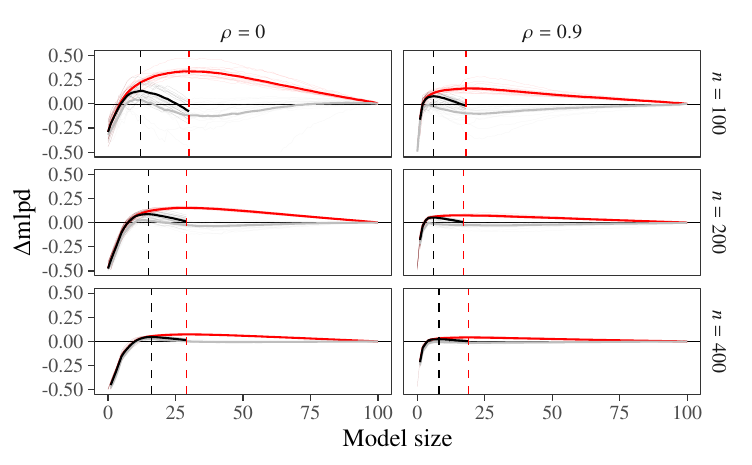}
         \caption{Using the R2D2 prior over regression coefficients.}
         \label{fig:r2d2-toy-example}
     \end{subfigure}
    \caption{Forward search simulation experiment. The \textcolor{red}{red} lines show the LOO-CV mlpd difference to the reference model over 20 iterations, the \textcolor{gray}{grey} lines show the test mlpd difference along those same paths, and the black show the corrected mlpd difference, produced using our correction from Section~\ref{sec:forward-order-stat}. The vertical dashed \textcolor{red}{red} and black lines indicate the point of maximum LOO-CV and corrected LOO-CV mlpd difference respectively.
    }
    \label{fig:forward-toy-example}
\end{figure}
\citet{gelman_prior_2017} reason that prior distributions in Bayesian inference should be considered generatively in terms of the potential observations consistent with the prior, and predictively in terms of how these observations align with the data we truly collect. \citet{cawley_over-fitting_2010} also discuss the benefits of using regularising techniques to avoid over-fitting in selection; generative priors are such a regulariser.

Joint shrinkage priors are a modern class of such generative priors which encode the statistician's belief on a predictive metric and filter this information through the model structure to define prior definitions on the model parameters. For instance, the R2D2(M2) prior \citep{zhang_bayesian_2022, aguilar_intuitive_2023} encodes a prior over the model's $R^2$ which is then mapped to a joint prior over the regression coefficients and the residual variance. By positing a lower expected predictive performance in terms of explained variance than what is implied by independent Gaussian priors, we shrink irrelevant regression coefficients more. This posterior shrinkage and reduced posterior uncertainty then naturally feeds through to the predictive distribution and thus the LOO-CV elpd estimate in Equation~\ref{eq:elpd-loo}. This means that candidate models make more similar predictions, the empirical distribution of elpd difference point estimates is more concentrated, and the magnitude of selection-induced bias is reduced. 
See Appendix~\ref{sec:r2d2} for more discussion on the specifics of the priors and the hyper-parameters used throughout this paper.

Repeating our example from Section~\ref{sec:toy-example}, but using a more regularising generative prior, we can reduce the degree of selection-induced bias along the search path. In Figure~\ref{fig:r2d2-toy-example}, we find that once more our heuristic identifies the correct model size after which no more predictors truly aid predictive performance, and recovers the shape of the test curve. We do, however, under-estimate the bias slightly in the $n = p$ regimes. This is likely due to the models making more similar predictions, and thus our proposed Equation~\ref{eq:order-stat-heuristic} will likely under-estimate the magnitude of the bias. However, the identified points of inflection are similar under both priors, reassuring us that our selection-induced bias correction is stable across different bias regimes.

Priors are not the only way one might consider reducing selection-induced bias in forward search; projection predictive inference is one information-theoretic approach to the task that has been proven to be very effective.
\subsection{Projection predictive inference}\label{sec:projpred}
We will later compare the model sizes selected following our bias correction to projection predictive inference. The projection approach is known to greatly reduce the variance in the relative performance estimates, leading to much smaller selection-induced bias \citep[see, e.g.][]{piironen_comparison_2017}.

Projection predictive inference was first introduced concretely by \citet{goutis_model_1998} and \citet{dupuis_variable_2003}, but follows the decision-theoretical idea presented previously by \citet{lindley_choice_1968}. In this paradigm, one begins with a large reference model one believes to be well-suited to prediction and subsequently constructs small models capable of approximately replicating its predictive performance; essentially, fitting sub-models to the fit of the reference model. 

We call the model we project the reference model onto the ``restricted model'' or ``sub-model'', and denote its parameter space by $\Theta^\perp$ (which may be completely unrelated to the parameter space of the reference model, $\Theta^\ast$). Further denote the predictor-conditional predictive distribution of the restricted model on $\tilde{n}$ unobserved data observations $\tilde{y} = (\tilde{y}_1,\dotsc,\tilde{y}_{\tilde{n}})^T$ by $q(\tilde{y}_i\mid\theta_\perp)$, for $\theta^\perp \in \Theta^\perp$ and $i \in \{1, \dotsc, n\}$, and that of the reference model by $p(\tilde{y}_i\mid\theta^\ast)$. Then we compute the projection draw-wise on $s \in \{1, \dotsc, S\}$ posterior draws from the reference model as
\begin{equation}
    \theta^\perp_s = \argmin_{\theta^\perp \in \Theta^\perp}
    \frac{1}{n} \sum_{i = 1}^n
    \mathbb{KL}\left\{\, p(\tilde{y}_i\mid\theta^\ast_s) \,\lVert\, q(\tilde{y}_i\mid\theta^\perp) \,\right\}. \label{eq:practical-projpred-orig}
\end{equation}
For implementation details see the papers by \citet{catalina_latent_2021, catalina_projection_2020}, \citet{weber_projection_2023}, and \citet{piironen_projective_2020}. \citet{mclatchie_advances_2024} outline further workflow steps to minimise the computational burden while achieving robust results.

Projective forward search begins by projecting the reference model onto the intercept-only model to establish the root of the search tree. 
Subsequently, the reference model is projected onto size-one models, from which we select the model whose posterior predictive distribution is most similar to the reference model's posterior predictive distribution by the KL divergence in line with the work of \citet{leamer_information_1979}. The search then continues as before. 
Because of the use of the KL divergence over LOO-CV, our order statistics-based bias estimation and correction can not be immediately translated to work with projection predictive inference. It is the disjoint between the KL divergence and the predictive performance noted by \citet{robert_projective_2014}, however, which is partially responsible for the reduced selection-induced bias in projection predictive inference.

As such, if the reference model (and sub-models) can be constructed in such a way that conforms with the requirements of projection predictive inference, then we recommend one begins with this. 
For observational families whose KL divergence can not be easily computed, or for model structures that can not be easily decomposed into additive components, cross-validation forward search (as described in the beginning of Section~\ref{sec:safe-forward-search}) can be a useful alternative.
\subsection{Real world data}\label{sec:real-world-data}
\begin{figure}[!t]
    \centering
    \includegraphics{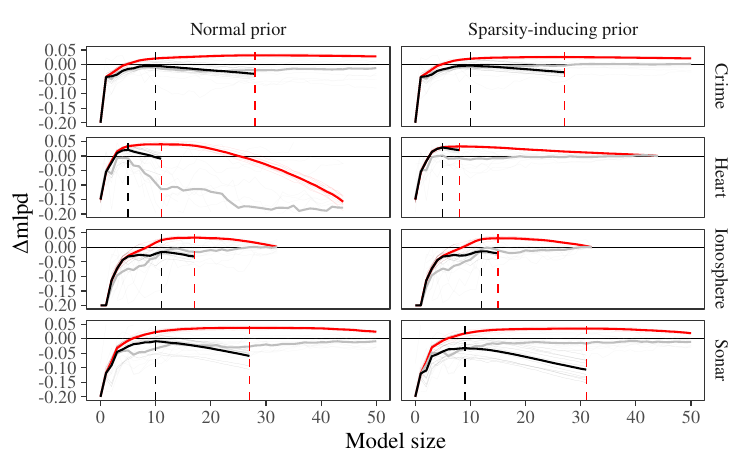}
    \caption{Real world datasets. The \textcolor{red}{red} lines show the LOO-CV mlpd difference to the reference model over CV folds, while the \textcolor{gray}{grey} lines show the test mlpd difference along those same folds. In black we see the corrected mlpd difference, produced using our correction from Section~\ref{sec:forward-order-stat}.}
    \label{fig:real-world-forward}
\end{figure}

To manifest the performance of our selection-induced bias correction beyond simulated settings, we compare it with classical stopping criteria, including projection predictive inference, across four publicly available real world datasets: one regression task \citep{redmond_data-driven_2002} and three binary classification \citep{gorman_analysis_1988, sigillito_classification_1989, cios_heart_96}
summarised in Table~\ref{tab:r2d2-hyperparams} (in Appendix~\ref{sec:r2d2}) 
after any necessary pre-processing.\footnote{The crime, ionosphere, and sonar data are freely available from the UCI machine learning benchmarks database: \url{https://archive.ics.uci.edu/ml/index.php}; the SPECT heart data are available from: \url{http://odds.cs.stonybrook.edu/}. Pre-processing is as described by \citet{piironen_comparison_2017}.} For the regression task we fit a linear regression reference model with a Gaussian observation family, and for the binary classification a Bernoulli generalised linear model as implemented in \pkg{brms} \citep{burkner_span_2017} using the probit link function. For each dataset, we perform forward search based on LOO-CV elpd, fitting sub-models with independent standard Gaussian priors over the regression coefficients in the first instance, and then again with sparse priors
(the R2D2 for regression tasks, and the regularised horseshoe for the classification tasks; see Appendix~\ref{sec:r2d2} for details on the hyper-parameters used for each task). 
We split all data into 10 cross-validation folds, leaving $10\%$ of the data for testing in each fold. 
In Figure~\ref{fig:real-world-forward} we show the cross-validated and test search paths for each of these, along with the corrected search paths as computed using Section~\ref{sec:forward-order-stat}.

Across all real data experiments, our corrected LOO-CV mlpd did not significantly exceed that of the reference model, except for the heart data. This indicates our correction effectively addresses selection-induced bias without falsely suggesting better performance than the reference model. In some cases, our correction appeared to under-correct for small model sizes but eventually recovered the true test mlpd. In the sonar data we witnessed the opposite: our proposal over-corrected bias. Even when our correction over- or under-estimates bias, it provides valuable insights into test mlpd behaviour.

We presently extend our selection-induced bias correction to produce a stopping criterion: we select the model size inducing the maximum \textit{corrected} mlpd along the search path as computed by Section~\ref{sec:forward-order-stat}. We compare this to simply stopping at the point of maximum LOO-CV (which we call the ``bulge'' rule), and projection predictive inference. 
Again, we do not look to identify the oracle model size, but rather we hope to select a model size having negligible excess loss compared to the oracle model. Alongside the bulge, we consider the so-called $2\sigma$ rule, which proposes that one should select the smallest model that is essentially indistinguishable from the model with the best LOO-CV performance, as judged by the two standard error simultaneous confidence interval. 

In projection predictive model selection, we first achieved a search path inline with the previous discussion in Section~\ref{sec:projpred}. We then choose the smallest model size $k$ whose (projected sub-model) elpd difference point estimate to the reference model (for which we use the full model fitted by standard Bayesian inference) is less than four.
This is in line with the recommendation of \citet[and discussed more in Appendix~\ref{sec:k2}]{mclatchie_advances_2024}.
We compare the model sizes chosen and their stability across forward search cross-validation folds in Figure~\ref{fig:real-world-select}. We make three primary observations:
\begin{figure}[!t]
    \centering
    \centering\includegraphics{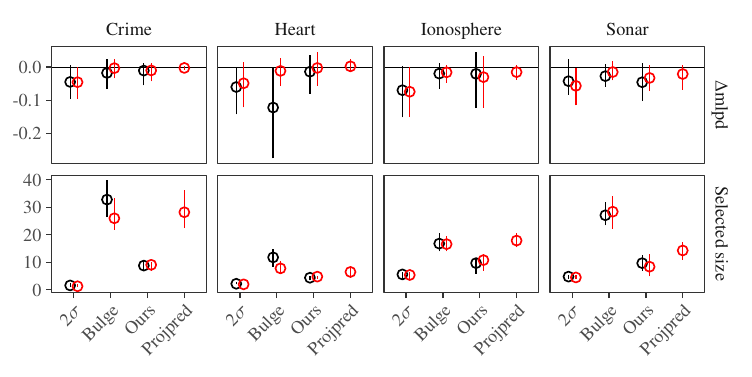}
    \caption{Real world datasets. In the bottom row we show the selected model sizes in four real world datasets across different heuristics when using standard independent Gaussian priors (in black), the sparsity-inducing priors (in \textcolor{red}{red}). The top row shows the test mlpd difference to the reference model of the model sizes selected by each heuristic.}
    \label{fig:real-world-select}
\end{figure}
\begin{enumerate}[nosep]
    \item projection predictive inference selects smaller model sizes than the most na\"ive approaches and does so stably;
    \item our correction achieves slightly smaller model sizes than the projective approach with similar mlpd, and requires fewer iterations of forward search than the bulge and $2\sigma$ rules (while achieving better mlpd than the latter two); and,
    \item regardless of model size selection heuristic, the use of shrinkage priors results in smaller and more stable model size selection.
\end{enumerate}
Across all data, the $2\sigma$ rule chooses the smallest sub-model sizes and the bulge rule chooses the largest. The model sizes selected by our correction are, by contrast, slightly smaller to those selected by projection predictive inference. There is more variation in the model size selected by the bulge rule. In fact, model size selection is generally more variable when using the independent Gaussian priors irrespective of the heuristic. Our correction is similarly stable to projection predictive, which is known to be very stable under the KL forward search heuristic, providing confidence in our own tool. Indeed, the good performance of projection predictive inference has been extensively covered by \citet{piironen_comparison_2017}. Our correction identifies models with better predictive performance than the more na\"ive $2\sigma$ and bulge rules, and with better stability.

The model size and mlpd difference estimated by our correction do not seem to be sensitive to the prior, unlike LOO-CV bulge model sizes; this is most prevalent in the heart data where the use of Gaussian priors results in vastly reduced predictive performance when using the bulge rule. 
Sparsity-inducing priors result in smaller and more stable model sizes being chosen across all data examples. This is likely related to our previous discussion on generative predictive priors: since they regularise in favour of predictive performance, they result in forward search paths that are less variable and less likely to introduce selection-induced bias (although may lead us to slightly under-estimate bias). Their predictive regularisation means that we are likely to more quickly and more stably identify predictive predictors, and ultimately identify more predictive predictor subsets.

The importance of identifying small models is particularly clear in this forward search example: when we must fit all models up to the bulge model size, this requires approximately $30^2 = 900$ model fits in the sonar and crime case studies. When these models are large, this can mean days of computation time. While the $2\sigma$ heuristic can identify small models with comparable predictive performance, it still requires fitting up to the bulge model size and then working back. In a word, it does not reduce the computational cost. As such, our bias correction is particularly attractive to those who are able to implement online stopping rules. 

It would also have been possible to construct a stopping rule based on the inclusion of zero in the two standard error range of the LOO-CV elpd \textit{difference} estimate. However, such heuristics struggle in the case of many weakly-relevant predictors. 
In particular, they are liable to identify too-early stopping points and achieve sub-optimal out-of-sample predictive performance when many predictors have weak (but non-zero) predictive information (e.g. in the ionosphere example). 
This is further discussed in Appendix~\ref{sec:alternative-heuristics}.

While our heuristic was not motivated from the perspective of model size selection, we find that even correcting just some of the selection-induced bias along forward search paths can reveal the behaviour of test performance. Using this corrected LOO-CV mlpd as a proxy for test mlpd is reasonable across these real data examples. As such, model size selection likely does not depend greatly on the assumed asymptotic correlation between LOO-CV and Bayes generalisation loss which we discussed in Section~\ref{sec:forward-order-stat}.
\section{Recommendations}
Cross-validation of predictive metrics is a common non-parametric method of making model selection decisions. It is particularly well-suited to cases when robust alternatives such as projection predictive inference are unavailable. Given the uncertainty present in LOO-CV estimates, it is possible that they are not always sufficient to make model selection decisions, or that they might encourage an incorrect decision based purely on noise. We discussed the risk of selecting between two candidate models (Scenario 1), presented a selection-induced bias estimation scheme based on order statistics in the case of selecting many candidate models (Scenario 2), and leveraged it to build a lightweight bias correction method in the forward search setting (Scenario 3).
\paragraph*{Scenario 1: the two model case}
In the case where we compare two models only, we can safely make a model selection decision if the elpd difference point estimate between them is greater than four in absolute value (see Appendix~\ref{sec:k2} for more discussion). 
Should the elpd difference point estimate be less than four in absolute value we have three options:
\begin{enumerate*}[label=(\arabic*), ref=\arabic*, nosep]
    \item if the models are not nested, combine them by model averaging or stacking;
    \item ensure the models' respective priors are reasonable and select the more complex of the two%
    ; or,
    \item keep them both as a set of best models.
\end{enumerate*}
This scenario is the least concerning of the three, and in most cases we can safely select the most predictive of the two provided its priors are reasonable.
\paragraph*{Scenario 2: the many model case}
In the many model case all previous recommendations apply, and we might further use our proposal from Section~\ref{sec:order-stat-proposal} to determine whether all models are equivalent in terms of predictive performance using order statistics. If we detect non-negligible selection-induced bias, then we do not make a model selection decision and instead perform model averaging or stacking (in the non-nested case). Our estimation procedure was shown to closely match empirical observations, and requires no further computation beyond the sufficient statistics of standard CV estimates. We further proposed a diagnostic to determine when the underlying assumptions of our proposition break down, and more computationally expensive approaches can be used instead.
\paragraph*{Scenario 3: forward search}
When the model space is large and we consider nested models, then we should first try projection predictive inference for model size selection. If the reference model is too difficult to fit, or if it belongs to an observation family which can not easily be implemented in the projective framework, then our lightweight selection-induced bias correction can have a direct application: by performing CV forward search and correcting the incremental elpd differences along the search path according to Section~\ref{sec:forward-order-stat}, we can recover an accurate estimate of the out-of-sample performance along the search path. And we do so at a significantly reduced cost to more computationally expensive procedures. In cases of severe over-fitting, and occasionally with shrinkage priors, it under-estimates the true bias slightly. However, we found that we did not need to correct the selection-induced bias perfectly to be able to accurately identify the point of predictive saturation in forward search. Indeed, just recovering the shape of the test elpd was sufficient for stable model size selection. Our procedure was found to effectively and reliably identify the point of over-fitting, and provides an accurate estimate of the bias, closely matching empirical observations.
\section{Discussion}\label{sec:discussion}
The use of order statistics in estimating selection-induced bias was introduced here in its most simple form, and is not without its limitations. We presently discuss those aspects which could be developed, and motivate the importance of future work in this area. 
\subsection{Limitations}
Naturally, our selection-induced bias correction hinges on a number of assumptions which may not always be met in practice. For instance, should models produce correlated predictions then the distributions the elpd point estimates are sampled from may not be independent leading to some error in the order statistic. We already witnessed this to a degree when using shrinkage priors. Likewise, the use of $\alpha = 0.5$ in the order statistic approximation is known to be conservative, although seemed to fit empirical data well. 

Further, while the elpd difference distribution of models with no extra predictive information from the baseline may theoretically be centred on zero, this is unlikely to be the case in practice. 
Indeed, this centering may in fact be slightly below zero. 
The choice of selection-induced bias correction based on the negative correlation results from \citet{watanabe_asymptotic_2010} is also heuristic, and may not be representative of the general case.
It is not immediately clear how the correction could be alternatively motivated in a robust manner, and we invite future work on the subject given the positive results we were able to achieve using a very simple model. 
Despite these limitations, our correction performed well across all simulated and real data exercise.
\subsection{Future directions}
It might be possible to improve on our simple correction by modelling the candidate models' LOO-CV elpd differences with a hierarchical model, and making decisions based on the posterior of the meta-model rather than the raw LOO-CV estimates \citep{gelman_why_2012,schmitt_meta_2023}. As well as trying to collapse all models into one group of in-differentiable models, as is the aim of partially-pooled models, one might produce a correlation-aware hierarchical model capable of identifying clusters of models exhibiting similar predictive performance. For example, one might consider a multivariate model with a similarity-based covariance matrix.

A more attainable extension is perhaps soft-thresholding the point at which we correct the LOO-CV elpd along the search path. The Fisher–Tippett–Gnedenko theorem tells us that the limiting distribution of properly-normalised maxima from a sequence of iid standard Gaussian observations follows a generalised extreme value distribution. Using this, or posterior weights from some meta-model, we might achieve some probability of having observed a truly predictive model from a set of candidates. This, as opposed to using the order statistic's mean as a hard threshold, would allow us to weight the correction we apply. While we did not observe any issue with the hard threshold we proposed, it is possible that in instances of many weakly-relevant predictors it is harder to separate between only negligibly relevant models, and our correction rule may be too rigid. 
\subsection*{Acknowledgments}
We acknowledge the computational resources provided by the Aalto Science-IT project, and the support of the Research Council of Finland Flagship programme: Finnish Center for Artificial Intelligence, and Research Council of Finland project ``Safe iterative model building'' (340721). We further thank Andrew R. Johnson, Frank Weber, Anna E. Riha, and Noa Kallionen for their helpful comments, and Leevi Lindgren for their implementation of forward search in \proglang{R}.
\bibliography{main}

\begin{thebibliography}{}

\bibitem[Aguilar and Bürkner, 2023]{aguilar_intuitive_2023}
Aguilar, J.~E. and Bürkner, P.-C. (2023).
\newblock Intuitive joint priors for {Bayesian} linear multilevel models: {The}
  {R2D2M2} prior.
\newblock {\em Electronic Journal of Statistics}, 17(1).

\bibitem[Ambroise and McLachlan, 2002]{ambroise_selection_2002}
Ambroise, C. and McLachlan, G.~J. (2002).
\newblock Selection bias in gene extraction on the basis of microarray
  gene-expression data.
\newblock {\em Proceedings of the National Academy of Sciences},
  99(10):6562--6566.

\bibitem[Arlot and Celisse, 2010]{arlot_survey_2010}
Arlot, S. and Celisse, A. (2010).
\newblock A survey of cross-validation procedures for model selection.
\newblock {\em Statistics surveys}, 4:40--79.

\bibitem[Barbieri and Berger, 2004]{barbieri_optimal_2004}
Barbieri, M.~M. and Berger, J.~O. (2004).
\newblock Optimal predictive model selection.
\newblock {\em The Annals of Statistics}, 32(3).

\bibitem[Bernardo and Smith, 1994]{bernardo_bayesian_1994}
Bernardo, J.~M. and Smith, A. F.~M. (1994).
\newblock {\em Bayesian Theory}.
\newblock John Wiley \& Sons.

\bibitem[Blom, 1960]{blom_statistical_1960}
Blom, G. (1960).
\newblock Statistical estimates and transformed beta-variables.
\newblock {\em Biometrika}, 47(1/2):210.

\bibitem[Brown et~al., 1998]{brown_multivariate_1998}
Brown, P.~J., Vannucci, M., and Fearn, T. (1998).
\newblock Multivariate {Bayesian} variable selection and prediction.
\newblock {\em Journal of the Royal Statistical Society Series B: Statistical
  Methodology}, 60(3):627--641.

\bibitem[Burnham and Anderson, 2002]{burnham_model_2002}
Burnham, K.~P. and Anderson, D.~R. (2002).
\newblock {\em Model {Selection} and {Multi}-{Model} {Inference}: {A}
  {Practical} {Information}-{Theoretic} {Approach}}.
\newblock Springer, 2nd edition.

\bibitem[Bürkner, 2017]{burkner_span_2017}
Bürkner, P.-C. (2017).
\newblock brms: {An} {R} package for {Bayesian} multilevel models using {Stan}.
\newblock {\em Journal of Statistical Software}, 80(1):1--28.
\newblock tex.encoding: UTF-8.

\bibitem[Bürkner et~al., 2020]{burkner_approximate_2020}
Bürkner, P.-C., Gabry, J., and Vehtari, A. (2020).
\newblock Approximate leave-future-out cross-validation for {Bayesian} time
  series models.
\newblock {\em Journal of Statistical Computation and Simulation},
  90(14):2499--2523.
\newblock arXiv:1902.06281 [stat].

\bibitem[Carvalho et~al., 2009]{carvalho_handling_2009}
Carvalho, C.~M., Polson, N.~G., and Scott, J.~G. (2009).
\newblock Handling sparsity via the horseshoe.
\newblock In van Dyk, D. and Welling, M., editors, {\em Proceedings of the
  Twelth International Conference on Artificial Intelligence and Statistics},
  volume~5 of {\em Proceedings of Machine Learning Research}, pages 73--80,
  Hilton Clearwater Beach Resort, Clearwater Beach, Florida USA. PMLR.

\bibitem[Catalina et~al., 2021]{catalina_latent_2021}
Catalina, A., Bürkner, P., and Vehtari, A. (2021).
\newblock Latent space projection predictive inference.
\newblock arXiv:2109.04702 [stat].

\bibitem[Catalina et~al., 2020]{catalina_projection_2020}
Catalina, A., Bürkner, P.-C., and Vehtari, A. (2020).
\newblock Projection predictive inference for generalized linear and additive
  multilevel models.
\newblock arXiv:2010.06994 [stat].

\bibitem[Cawley and Talbot, 2010]{cawley_over-fitting_2010}
Cawley, G.~C. and Talbot, N. L.~C. (2010).
\newblock On over-fitting in model selection and subsequent selection bias in
  performance evaluation.
\newblock {\em Journal of Machine Learning Research}, 11:2079--2107.

\bibitem[Cooper et~al., 2023]{cooper_cross-validatory_2023}
Cooper, A., Simpson, D., Kennedy, L., Forbes, C., and Vehtari, A. (2023).
\newblock Cross-validatory model selection for {Bayesian} autoregressions with
  exogenous regressors.
\newblock arXiv:2301.08276 [stat].

\bibitem[Dupuis and Robert, 2003]{dupuis_variable_2003}
Dupuis, J.~A. and Robert, C.~P. (2003).
\newblock Variable selection in qualitative models via an entropic explanatory
  power.
\newblock {\em Journal of Statistical Planning and Inference}, 111(1-2):77--94.

\bibitem[Efron and Tibshirani, 1993]{efron_introduction_1993}
Efron, B. and Tibshirani, R. (1993).
\newblock {\em An introduction to the bootstrap}.
\newblock Number~57 in Monographs on statistics and applied probability.
  Chapman \& Hall, New York.

\bibitem[Galatenko et~al., 2015]{galatenko_highly_2015}
Galatenko, V.~V., Shkurnikov, M.~Y., Samatov, T.~R., Galatenko, A.~V.,
  Mityakina, I.~A., Kaprin, A.~D., Schumacher, U., and Tonevitsky, A.~G.
  (2015).
\newblock Highly informative marker sets consisting of genes with low
  individual degree of differential expression.
\newblock {\em Scientific Reports}, 5(1):14967.

\bibitem[Garthwaite and Mubwandarikwa, 2010]{garthwaite_selection_2010}
Garthwaite, P.~H. and Mubwandarikwa, E. (2010).
\newblock Selection of weights for weighted model averaging: Prior weights for
  weighted model averaging.
\newblock {\em Australian \& New Zealand Journal of Statistics},
  52(4):363--382.

\bibitem[Geisser, 1975]{Geisser:1975}
Geisser, S. (1975).
\newblock The predictive sample reuse method with applications.
\newblock {\em Journal of the {American} Statistical Association},
  70(350):320--328.

\bibitem[Geisser and Eddy, 1979]{geisser_eddy_predictive_1979}
Geisser, S. and Eddy, W.~F. (1979).
\newblock A predictive approach to model selection.
\newblock {\em Journal of the American Statistical Association},
  74(365):153--160.

\bibitem[Gelfand and Ghosh, 1998]{gelfand_model_1998}
Gelfand, A. and Ghosh, S.~K. (1998).
\newblock Model choice: a minimum posterior predictive loss approach.
\newblock {\em Biometrika}, 85(1):1--11.

\bibitem[Gelfand, 1996]{gelfand_model_1996}
Gelfand, A.~E. (1996).
\newblock Model determination using sampling-based methods.
\newblock {\em Markov chain Monte Carlo in practice}, 4:145--161.
\newblock Publisher: London.

\bibitem[Gelfand et~al., 1992]{gelfand_model_1992}
Gelfand, A.~E., Dey, D.~K., and Chang, H. (1992).
\newblock Model determination using predictive distributions with
  implementation via sampling-based methods.
\newblock Technical report, Stanford Univ CA Dept of Statistics.

\bibitem[Gelman, 2022]{gelman_im_2022}
Gelman, A. (2022).
\newblock I’m skeptical of that claim that “{Cash} aid to poor mothers
  increases brain activity in babies”.

\bibitem[Gelman et~al., 2013]{gelman_bayesian_2014}
Gelman, A., Carlin, J.~B., Stern, H.~S., Dunson, D.~B., Vehtari, A., and Rubin,
  D.~B. (2013).
\newblock {\em Bayesian {Data} {Analysis}}.
\newblock Chapman and Hall/CRC, 3rd edition.

\bibitem[Gelman et~al., 2012]{gelman_why_2012}
Gelman, A., Hill, J., and Yajima, M. (2012).
\newblock Why we (usually) don't have to worry about multiple comparisons.
\newblock {\em Journal of Research on Educational Effectiveness},
  5(2):189--211.
\newblock Publisher: Routledge \_eprint:
  https://doi.org/10.1080/19345747.2011.618213.

\bibitem[Gelman et~al., 2014]{gelman_understanding_2014}
Gelman, A., Hwang, J., and Vehtari, A. (2014).
\newblock Understanding predictive information criteria for {Bayesian} models.
\newblock {\em Statistics and Computing}, 24(6):997--1016.

\bibitem[Gelman et~al., 2017]{gelman_prior_2017}
Gelman, A., Simpson, D., and Betancourt, M. (2017).
\newblock The prior can often only be understood in the context of the
  likelihood.
\newblock {\em Entropy}, 19(10):555.

\bibitem[Gelman et~al., 1996]{gelman_posterior_1996}
Gelman, A., Xiao-Li, M., and Stern, H.~S. (1996).
\newblock Posterior predictive assessment of model fitness via realized
  discrepancies.
\newblock {\em Statistica Sinica}, 6(4):733--760.

\bibitem[George and McCulloch, 1993]{george_variable_1993}
George, E.~I. and McCulloch, R.~E. (1993).
\newblock Variable selection via {Gibbs} sampling.
\newblock {\em Journal of the American Statistical Association},
  88(423):881--889.

\bibitem[Gorman and Sejnowski, 1988]{gorman_analysis_1988}
Gorman, R. and Sejnowski, T.~J. (1988).
\newblock Analysis of hidden units in a layered network trained to classify
  sonar targets.
\newblock {\em Neural Networks}, 1(1):75--89.

\bibitem[Goutis, 1998]{goutis_model_1998}
Goutis, C. (1998).
\newblock Model choice in generalised linear models: a {Bayesian} approach via
  {Kullback}-{Leibler} projections.
\newblock {\em Biometrika}, 85(1):29--37.

\bibitem[Han and Carlin, 2001]{han_markov_2001}
Han, C. and Carlin, B.~P. (2001).
\newblock Markov chain {Monte} {Carlo} methods for computing {Bayes} factors: A
  comparative review.
\newblock {\em Journal of the American Statistical Association},
  96(455):1122--1132.

\bibitem[Harrell, 2001]{harrell_regression_2001}
Harrell, F.~E. (2001).
\newblock {\em Regression Modeling Strategies: With Applications to Linear
  Models, Logistic Regression, and Survival Analysis}.
\newblock Springer {Series} in {Statistics}. Springer New York, New York, NY.

\bibitem[Harter, 1961]{harter_expected_1961}
Harter, H.~L. (1961).
\newblock Expected values of normal order statistics.
\newblock {\em Biometrika}, 48(1/2):151.

\bibitem[Hoeting et~al., 1999]{hoeting_bayesian_1999}
Hoeting, J.~A., Madigan, D., Raftery, A.~E., and Volinsky, C.~T. (1999).
\newblock Bayesian model averaging: a tutorial (with comments by {M}. {Clyde},
  {David} {Draper} and {E}. {I}. {George}, and a rejoinder by the authors.
\newblock {\em Statistical Science}, 14(4).

\bibitem[Jeffreys, 1998]{jeffreys_theory_1998}
Jeffreys, H. (1998).
\newblock {\em Theory of probability}.
\newblock Oxford classic texts in the physical sciences. Clarendon Press ;
  Oxford University Press, Oxford [Oxfordshire] : New York, 3rd ed edition.

\bibitem[Kass and Raftery, 1995]{kass_bayes_1995}
Kass, R.~E. and Raftery, A.~E. (1995).
\newblock Bayes factors.
\newblock {\em Journal of the American Statistical Association},
  90(430):773--795.

\bibitem[Key et~al., 1999]{key_bayesian_1999}
Key, J., Pericchi, L., and Smith, A. F.~M. (1999).
\newblock Bayesian model choice: What and why?
\newblock {\em Bayesian Statistics}.

\bibitem[Krzysztof~Cios, 2001]{cios_heart_96}
Krzysztof~Cios, Lukasz~Kurgan, L.~G. (2001).
\newblock {SPECTF} heart data.
\newblock UCI Machine Learning Repository.
\newblock {DOI}: https://doi.org/10.24432/C5N015.

\bibitem[Laud and Ibrahim, 1995]{laud_predictive_1995}
Laud, P.~W. and Ibrahim, J.~G. (1995).
\newblock Predictive model selection.
\newblock {\em Journal of the Royal Statistical Society: Series B
  (Methodological)}, 57(1):247--262.

\bibitem[Le and Clarke, 2022]{le_model_2022}
Le, T.~M. and Clarke, B.~S. (2022).
\newblock Model averaging is asymptotically better than model selection for
  prediction.
\newblock {\em Journal of Machine Learning Research}, 23(33):1--53.

\bibitem[Leamer, 1979]{leamer_information_1979}
Leamer, E.~E. (1979).
\newblock Information criteria for choice of regression models: A comment.
\newblock {\em Econometrica}, 47(2):507.

\bibitem[Lindley, 1968]{lindley_choice_1968}
Lindley, D.~V. (1968).
\newblock The choice of variables in multiple regression.
\newblock {\em Journal of the Royal Statistical Society: Series B
  (Methodological)}, 30(1):31--53.

\bibitem[Marriott et~al., 2001]{marriott_bayesian_2001}
Marriott, J.~M., Spencer, N.~M., and Pettitt, A.~N. (2001).
\newblock A {Bayesian} approach to selecting covariates for prediction.
\newblock {\em Scandinavian Journal of Statistics}, 28(1):87--97.

\bibitem[McLatchie et~al., 2024]{mclatchie_advances_2024}
McLatchie, Y., Rögnvaldsson, S., Weber, F., and Vehtari, A. (2024).
\newblock Advances in projection predictive inference.
\newblock arXiv:2306.15581 [stat].

\bibitem[Merkle et~al., 2019]{merkle_bayesian_2019}
Merkle, E.~C., Furr, D., and Rabe-Hesketh, S. (2019).
\newblock Bayesian {Comparison} of {Latent} {Variable} {Models}: {Conditional}
  {Versus} {Marginal} {Likelihoods}.
\newblock {\em Psychometrika}, 84(3):802--829.

\bibitem[Narisetty and He, 2014]{narisetty_bayesian_2014}
Narisetty, N.~N. and He, X. (2014).
\newblock Bayesian variable selection with shrinking and diffusing priors.
\newblock {\em The Annals of Statistics}, 42(2).

\bibitem[Nilsson, 1998]{nilsson_artificial_1998}
Nilsson, N.~J. (1998).
\newblock {\em Artificial Intelligence: A New Synthesis}.
\newblock Elsevier.

\bibitem[Ntzoufras et~al., 2000]{ntzoufras_stochastic_2000}
Ntzoufras, I., Forster, J.~J., and Dellaportas, P. (2000).
\newblock Stochastic search variable selection for log-linear models.
\newblock {\em Journal of Statistical Computation and Simulation},
  68(1):23--37.

\bibitem[Oelrich et~al., 2020]{oelrich_when_2020}
Oelrich, O., Ding, S., Magnusson, M., Vehtari, A., and Villani, M. (2020).
\newblock When are {Bayesian} model probabilities overconfident?
\newblock arXiv:2003.04026 [math, stat].

\bibitem[O'Hara and Sillanpää, 2009]{ohara_review_2009}
O'Hara, R.~B. and Sillanpää, M.~J. (2009).
\newblock A review of {Bayesian} variable selection methods: what, how and
  which.
\newblock {\em Bayesian Analysis}, 4(1).

\bibitem[Piironen et~al., 2020]{piironen_projective_2020}
Piironen, J., Paasiniemi, M., and Vehtari, A. (2020).
\newblock Projective inference in high-dimensional problems: Prediction and
  feature selection.
\newblock {\em Electronic Journal of Statistics}, 14(1).
\newblock arXiv:1810.02406 [cs, stat].

\bibitem[Piironen and Vehtari, 2017a]{piironen_comparison_2017}
Piironen, J. and Vehtari, A. (2017a).
\newblock Comparison of {Bayesian} predictive methods for model selection.
\newblock {\em Statistics and Computing}, 27(3):711--735.

\bibitem[Piironen and Vehtari, 2017b]{piironen_sparsity_2017}
Piironen, J. and Vehtari, A. (2017b).
\newblock Sparsity information and regularization in the horseshoe and other
  shrinkage priors.
\newblock {\em Electronic Journal of Statistics}, 11(2).

\bibitem[Raftery and Zheng, 2003]{raftery_discussion_2003}
Raftery, A.~E. and Zheng, Y. (2003).
\newblock Discussion: Performance of {Bayesian} model averaging.
\newblock {\em Journal of the American Statistical Association},
  98(464):931--938.

\bibitem[Redmond and Baveja, 2002]{redmond_data-driven_2002}
Redmond, M. and Baveja, A. (2002).
\newblock A data-driven software tool for enabling cooperative information
  sharing among police departments.
\newblock {\em European Journal of Operational Research}, 141(3):660--678.

\bibitem[Reunanen, 2003]{reunanen_overfitting_2003}
Reunanen, J. (2003).
\newblock Overfitting in making comparisons between variable selection methods.
\newblock {\em Journal of Machine Learning Research}, 3:1371--1382.

\bibitem[Robert, 2014]{robert_projective_2014}
Robert, C. (2014).
\newblock Projective covariate selection.

\bibitem[Royston, 1982]{royston_algorithm_1982}
Royston, J.~P. (1982).
\newblock Algorithm {AS} 177: Expected normal order statistics (exact and
  approximate).
\newblock {\em Applied Statistics}, 31(2):161.

\bibitem[Scarrott and MacDonald, 2012]{scarrott_review_2012}
Scarrott, C. and MacDonald, A. (2012).
\newblock Review of extreme value threshold estimation and uncertainty
  quantification.
\newblock {\em REVSTAT-Statistical Journal}, pages 33--60 Pages.
\newblock Artwork Size: 33–60 Pages Publisher: REVSTAT-Statistical Journal.

\bibitem[Schmitt et~al., 2023]{schmitt_meta_2023}
Schmitt, M., Radev, S.~T., and B\"urkner, P.-C. (2023).
\newblock Meta-uncertainty in bayesian model comparison.
\newblock In Ruiz, F., Dy, J., and van~de Meent, J.-W., editors, {\em
  Proceedings of The 26th International Conference on Artificial Intelligence
  and Statistics}, volume 206 of {\em Proceedings of Machine Learning
  Research}, pages 11--29. PMLR.

\bibitem[Scholz and Bürkner, 2022]{scholz_prediction_2022}
Scholz, M. and Bürkner, P.-C. (2022).
\newblock Prediction can be safely used as a proxy for explanation in causally
  consistent {Bayesian} generalized linear models.
\newblock arXiv:2210.06927 [stat].

\bibitem[Shao, 1993]{shao_linear_1993}
Shao, J. (1993).
\newblock Linear model selection by cross-validation.
\newblock {\em Journal of the American Statistical Association},
  88(422):486--494.

\bibitem[Sigillito et~al., 1989]{sigillito_classification_1989}
Sigillito, V., Wing, S., Hutton, L.~V., and Baker, K. (1989).
\newblock Classification of radar returns from the ionosphere using neural
  networks.
\newblock {\em Johns Hopkins APL Technical Digest}, 10:262--266.

\bibitem[Sivula et~al., 2022]{sivula_uncertainty_2022}
Sivula, T., Magnusson, M., Matamoros, A.~A., and Vehtari, A. (2022).
\newblock Uncertainty in {Bayesian} leave-one-out cross-validation based model
  comparison.
\newblock arXiv:2008.10296 [stat].

\bibitem[Spiegelhalter et~al., 2002]{spiegelhalter_bayesian_2002}
Spiegelhalter, D.~J., Best, N.~G., Carlin, B.~P., and van~der Linde, A. (2002).
\newblock Bayesian measures of model complexity and fit.
\newblock {\em Journal of the Royal Statistical Society: Series B (Statistical
  Methodology)}, 64(4):583--639.

\bibitem[Stone, 1974]{stone_cross-validatory_1974}
Stone, M. (1974).
\newblock Cross-validatory choice and assessment of statistical predictions.
\newblock {\em Journal of the Royal Statistical Society: Series B
  (Methodological)}, 36(2):111--133.

\bibitem[Tibshirani and Tibshirani, 2009]{tibshirani_bias_2009}
Tibshirani, R.~J. and Tibshirani, R. (2009).
\newblock A bias correction for the minimum error rate in cross-validation.
\newblock {\em The Annals of Applied Statistics}, 3(2).

\bibitem[Troller-Renfree et~al., 2022]{troller-renfree_impact_2022}
Troller-Renfree, S.~V., Costanzo, M.~A., Duncan, G.~J., Magnuson, K.,
  Gennetian, L.~A., Yoshikawa, H., Halpern-Meekin, S., Fox, N.~A., and Noble,
  K.~G. (2022).
\newblock The impact of a poverty reduction intervention on infant brain
  activity.
\newblock {\em Proceedings of the National Academy of Sciences},
  119(5):e2115649119.

\bibitem[Vehtari et~al., 2023]{vehtari_loo_2023}
Vehtari, A., Gabry, J., Magnusson, M., Yao, Y., Bürkner, P.-C., Paananen, T.,
  and Gelman, A. (2023).
\newblock loo: Efficient leave-one-out cross-validation and waic for bayesian
  models.
\newblock R package version 2.6.0.

\bibitem[Vehtari et~al., 2017]{vehtari_practical_2017}
Vehtari, A., Gelman, A., and Gabry, J. (2017).
\newblock Practical {Bayesian} model evaluation using leave-one-out
  cross-validation and {WAIC}.
\newblock {\em Statistics and Computing}, 27(5):1413--1432.

\bibitem[Vehtari and Lampinen, 2002]{vehtari_bayesian_2002}
Vehtari, A. and Lampinen, J. (2002).
\newblock Bayesian model assessment and comparison using cross-validation
  predictive densities.
\newblock {\em Neural Computation}, 14(10):2439--2468.

\bibitem[Vehtari and Ojanen, 2012]{vehtari_survey_2012}
Vehtari, A. and Ojanen, J. (2012).
\newblock A survey of {Bayesian} predictive methods for model assessment,
  selection and comparison.
\newblock {\em Statistics Surveys}, 6(none).

\bibitem[Vehtari et~al., 2022]{vehtari_pareto_2022}
Vehtari, A., Simpson, D., Gelman, A., Yao, Y., and Gabry, J. (2022).
\newblock Pareto {Smoothed} {Importance} {Sampling}.
\newblock arXiv:1507.02646 [stat].

\bibitem[Wang and Gelman, 2015]{wang_difficulty_2015}
Wang, W. and Gelman, A. (2015).
\newblock Difficulty of selecting among multilevel models using predictive
  accuracy.
\newblock {\em Statistics and Its Interface}, 8(2):153--160.

\bibitem[Watanabe, 2010]{watanabe_asymptotic_2010}
Watanabe, S. (2010).
\newblock Asymptotic equivalence of {Bayes} cross validation and widely
  applicable information criterion in singular learning theory.
\newblock arXiv:1004.2316 [cs].

\bibitem[Watanabe, 2013]{watanabe_widely_2013}
Watanabe, S. (2013).
\newblock A widely applicable {Bayesian} information criterion.
\newblock {\em Journal of Machine Learning Research}, page~31.

\bibitem[Weber and Vehtari, 2023]{weber_projection_2023}
Weber, F. and Vehtari, A. (2023).
\newblock Projection predictive variable selection for discrete response
  families with finite support.
\newblock arXiv:2301.01660 [stat].

\bibitem[Yanchenko et~al., 2023a]{yanchenko_space_2023}
Yanchenko, E., Bondell, H.~D., and Reich, B.~J. (2023a).
\newblock {R2D2} goes to space! {A} principled approach to setting prior
  distributions on spatial parameters.
\newblock arXiv:2301.09951 [stat].

\bibitem[Yanchenko et~al., 2023b]{yanchenko_r2d2_2023}
Yanchenko, E., Bondell, H.~D., and Reich, B.~J. (2023b).
\newblock The {R2D2} prior for generalized linear mixed models.
\newblock arXiv:2111.10718 [stat].

\bibitem[Yao et~al., 2023]{yao_locking_2023}
Yao, Y., Carvalho, L.~M., Mesquita, D., and McLatchie, Y. (2023).
\newblock Locking and quacking: Stacking {Bayesian} model predictions by
  log-pooling and superposition.
\newblock arXiv:2305.07334 [cs, stat].

\bibitem[Yao et~al., 2021]{yao_bayesian_2021}
Yao, Y., Pirš, G., Vehtari, A., and Gelman, A. (2021).
\newblock Bayesian hierarchical stacking: Some models are (somewhere) useful.
\newblock {\em Bayesian Analysis}, -1(-1).
\newblock arXiv:2101.08954 [cs, stat].

\bibitem[Yao et~al., 2018]{yao_using_2018}
Yao, Y., Vehtari, A., Simpson, D., and Gelman, A. (2018).
\newblock Using stacking to average {Bayesian} predictive distributions (with
  discussion).
\newblock {\em Bayesian Analysis}, 13(3):917--1007.

\bibitem[Yates et~al., 2021]{yates_parsimonious_2021}
Yates, L.~A., Richards, S.~A., and Brook, B.~W. (2021).
\newblock Parsimonious model selection using information theory: a modified
  selection rule.
\newblock {\em Ecology}, 102(10).

\bibitem[Zhang and Stephens, 2009]{zhang_new_2009}
Zhang, J. and Stephens, M.~A. (2009).
\newblock A new and efficient estimation method for the generalized {Pareto}
  distribution.
\newblock {\em Technometrics}, 51(3):316--325.

\bibitem[Zhang et~al., 2022]{zhang_bayesian_2022}
Zhang, Y.~D., Naughton, B.~P., Bondell, H.~D., and Reich, B.~J. (2022).
\newblock Bayesian regression using a prior on the model fit: The {R2}-{D2}
  shrinkage prior.
\newblock {\em Journal of the American Statistical Association},
  117(538):862--874.

\end{thebibliography}
\newpage
%
\counterwithin*{equation}{section}
\counterwithin*{figure}{section}
\renewcommand\theequation{\thesection.\arabic{equation}}
\renewcommand\thefigure{\thesection.\arabic{figure}}
\begin{appendices} 
\section{Connection between elpd difference and model averaging}\label{sec:k2}
Choosing a model based on elpd point estimate, as we have proposed throughout this paper, has a deeper connection to Bayesian model averaging. Suppose we are in the $K = 2$ case, we might say that one model is sufficiently better than another to feel safe making a model selection decision if
\begin{equation}\label{eq:delta-heuristic}
    \left\vert\Delta\elpdHat{\Ma,\Mb}{y}\right\vert < 4.
\end{equation}
This was initially considered by \citet{sivula_uncertainty_2022} in terms of the reliability of the standard error estimate: when the elpd difference point estimate is this low, it is likely that the elpd difference \textit{standard error} estimate is too small. \citet{mclatchie_advances_2024} later used this flavour of selection heuristic in the context of projection predictive inference.

\begin{figure}[!t]
    \centering
    \includegraphics{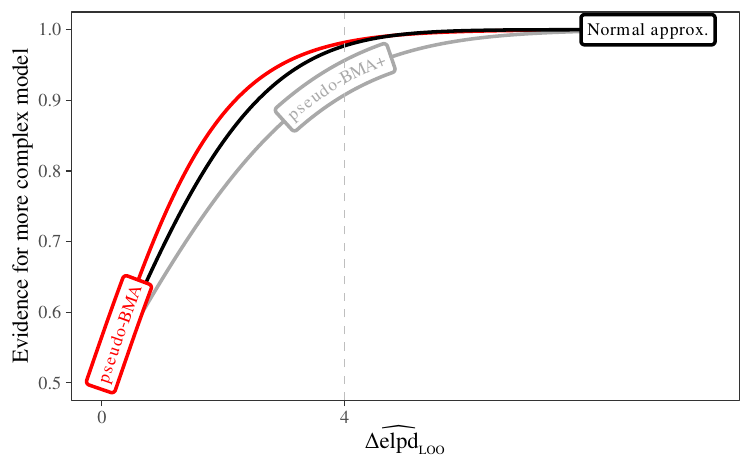}
    \caption{Theoretical connection between elpd difference and BMA. The relationship between pseudo-BMA weights (Equation~\ref{eq:pseudo-bma}), the pseudo-BMA+ weights (Equation~\ref{eq:pseudo-bma+}), and the probability of one model being better than another based on the normal approximation of the elpd difference distribution (Equation~\ref{eq:normal-approx-diff}). We fix the standard error of the elpd difference normal approximation to be $\mathrm{se}(\Delta\elpdHatPlain) = 2$ for demonstration purposes. For this choice of standard error, Equation~\ref{eq:normal-approx-diff} acts very similarly to the probit function (naturally, as it is equivalent to the Gaussian cumulative density function computed up to zero). This probit is precisely the pseudo-BMA weight of the more complex model, and mildly altered becomes the pseudo-BMA+ weights. At the point $\Delta\elpdHatPlain\approx4$, and the probability of it being better than the less complex model is approximately $98\%$.}
    \label{fig:elpd-diff-pmba}
\end{figure}

Suppose we have once more the situation where we have two models $\Ma$ and $\Mb$. If we consider a normal approximation of the LOO-CV elpd difference estimate, then the probability that the more predictive model, say $\Ma$, is truly better than the other is 
\begin{equation}\label{eq:normal-approx-diff}
    \prob{\elpd{\Ma}{y} \geq \elpd{\Mb}{y}} = \int_{-\infty}^0 \normal\left(z\mid\Delta\elpdHatPlain,\mathrm{se}\left(\Delta\elpdHatPlain\right)\right) \,\mathrm{d}z.
\end{equation}
Suppose the standard error is two for demonstration purposes, then when the point estimate is $4$, the probability of one model being better than another is approximately $98\%$. This is shown in Figure~\ref{fig:elpd-diff-pmba}.

Likewise, \citet{yao_bayesian_2021} present an AIC-type model weighting score which uses LOO-CV predictive densities which they call pseudo-Bayesian model averaging (pseudo-BMA) weights. The weight of the more predictive model of the two compared is
\begin{equation}\label{eq:pseudo-bma}
    w_{k} = \frac{1}{1 + \exp\left(-\Delta\elpdHatPlain\right)}.
\end{equation}
Once more, we find that should we have two models with an elpd difference point estimate of 4, then the psuedo-BMA weight of the more predictive model is again approximately $0.98$. This weighting system, however, ignores the uncertainty associated with the point estimate. To remedy this, \citet{yao_bayesian_2021} present what they call pseudo-BMA+ weights that integrate over this uncertainty, where the weight of the more predictive model is now
\begin{equation}\label{eq:pseudo-bma+}
    w_{k}^+ = \int_{-\infty}^\infty \frac{\normal\left(z\mid0,\mathrm{se}\left(\Delta\elpdHatPlain\right)\right)}{1 + \exp\left(-\Delta\elpdHatPlain - z\right)} \,\mathrm{d}z.
\end{equation}
Assuming a standard error of two as before, the weight of the more predictive model is again over $90\%$, albeit lower than with pseudo-BMA and normal approximation calculations. This is again shown in Figure~\ref{fig:elpd-diff-pmba}. \cite{yao_bayesian_2021} show that their pseudo-BMA+ weightings perform better than standard pseudo-BMA, and even with this stricter measurement, looking at just the point estimate of the elpd difference is safe when making model selection decisions in the two-model case. 

Now we can reflect selection-induced bias in terms of the value 4: if the bias is big enough to impact a decision made based on this \textit{ad hoc} rule-of-four, then we may start to falsely give some models zero weight in pseudo-BMA. For instance, if the bias is much less than $4$, then we could say it is negligible. But if the bias grows above $4$ then we have problems using the elpd estimates for model averaging. We do not, however, propose this number as a hard threshold.
\section{Empirical demonstration of order statistics}\label{sec:order-stat-demo}
We presently consider data from a study on the effect of poverty reduction on infant brain activity \citep{troller-renfree_impact_2022}. In this study, the authors report the impact of a cash incentive to mothers on the brain activity in the first year of life of their children. The mothers were randomised to receive an monthly, unconditional cash grant of either ``significant'' or ``nominal'' size. The children's brain activities were then measured by resting electroencephalography (EEG) after a year to ascertain whether these cash transfers changed infant brain activity. We build a collection of different models on the ``alpha brain band'' data, varying the number of predictors, the observation family, and the data pre-processing.

In particular, we create a pre-treatment score similarly to \citet{gelman_im_2022} by combining birth weight, gestational age, the mother’s health, and amount of smoking and drinking during pregnancy. These predictors are scaled to have zero mean and unit standard deviation, and the signs of smoking and drinking predictors are switched before being summed. We construct generalised linear models (GLMs) with just the treatment effect in a first instance, before adding the pre-treatment score, household income, birth weight, and mother's race incrementally. We then also investigate GLMs with a Gaussian observation family, and a log-normal observation family. Having fit these 10 models, we then identify the median model in terms of LOO-CV elpd point estimate, and use this as the baseline model.

\begin{figure}
    \centering
    \includegraphics{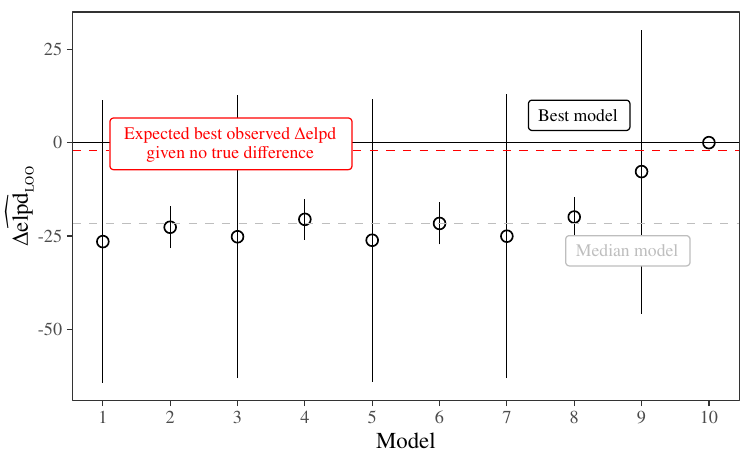}
    \caption{EEG cash incentive data. We show in the point ranges the LOO-CV elpd differences and difference standard errors of $10$ models with respect to the baseline model (chosen as the model with the median LOO-CV elpd, and located at index $6$). We plot the expected highest elpd difference under the assumption that no model is better than the baseline with order statistics in the \textcolor{red}{red} dashed line. This identifies at least one of the models (the tenth) to be significantly better than the baseline model.}
    \label{fig:order-stat-demo}
\end{figure}

In Figure~\ref{fig:order-stat-demo} we show the LOO-CV elpd differences of the remaining nine models to the baseline (shown in the plain dot at index $6$). We compute the order statistic-based bias estimate, $S^{(K)}\hat{\sigma}_K$ with $K = 9$, as previously described and plot this next to the model LOO-CV elpd differences. These results would have us believe that at least one of the models (specifically the tenth) is significantly better than the baseline, and that in this case we can safely select it. The ninth model lies very close to the border marked out the order statistic; if we were concerned about selection-induced bias in this case, we could average or stack these two seemingly more predictive models rather than select only one.
\section{Bias estimation under different assumed asymptotic cross-validated correlation regimes}\label{sec:alt-correlations}
As previously mentioned, the origin of our bias correction lies in the work of \citet[Table 3]{watanabe_asymptotic_2010}. They discuss a notable negative correlation between cross-validated scores and average Bayes generalisation loss, observed in just one data instance. This arises due to excluding an influential observation from training, leading to a skewed posterior predictive distribution and the observation's seemingly anomalous appearance in the test set. The correction of LOO-CV elpd estimates thus necessitates a negatively-correlated correction. In essence, the adjusted elpd difference should be mirrored across the median of candidate differences. The choice of magnitude $1.5$ in our bias correction reflects a negative correlation around $-0.8$ noted by \citet{watanabe_asymptotic_2010}, albeit from a single data instance. Recognising this limitation, we repeat the correction in the simulated forward search experiments with values $\{1, 1.5, 2\}$, corresponding to correlations $\{0, -0.5, -1\}$.

\begin{figure}[!t]
     \centering
     \begin{subfigure}{\textwidth}
         \centering
         \includegraphics{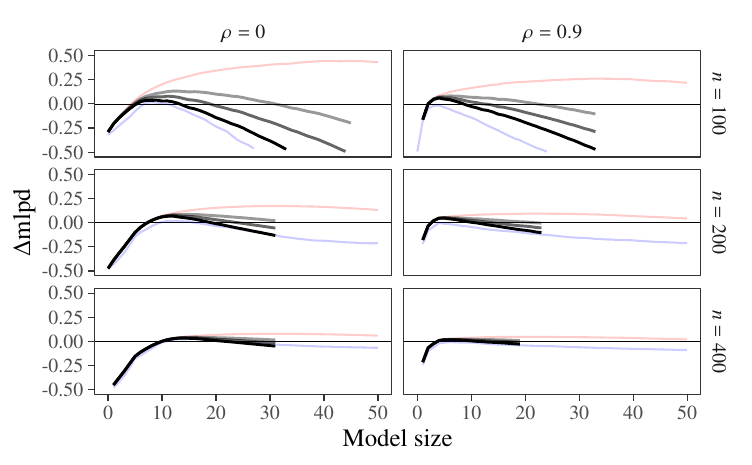}
         \caption{Independent Gaussian priors over regression coefficients.}
         \label{fig:alt-normal-toy-example}
     \end{subfigure}
     \hfill
     \begin{subfigure}{\textwidth}
         \centering
         \includegraphics{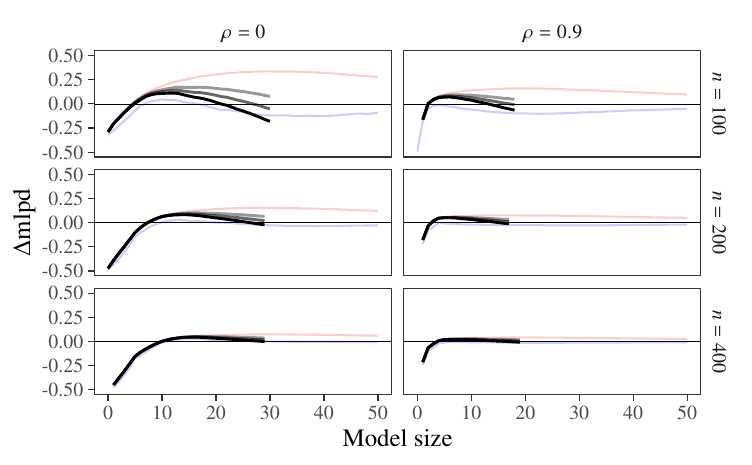}
         \caption{R2D2 prior over regression coefficients.}
         \label{fig:alt-r2d2-toy-example}
     \end{subfigure}
    \caption{Forward search simulation experiment across different priors, and under different assumed asymptotic correlations between LOO-CV estimate and average Bayes generalisation loss. The \textcolor{red}{red} line show the LOO-CV mlpd difference to the reference model, and the \textcolor{blue}{blue} lines show the test mlpd difference. In black we show the mlpd difference corrected according to our proposal with assumed correlation $-1$, in \textcolor{darkgray}{lighter grey} we assume correlation $-0.5$, and in the \textcolor{gray}{lightest grey} an assumed $0$ correlation. Higher assumed correlation results in more correction while assuming, for example, a zero correlation under-estimates the selection-induced in most cases.}
    \label{fig:alt-forward-toy-example}
\end{figure}

In Figure~\ref{fig:alt-forward-toy-example} we see how these different values perform across different $n$ and $\rho$ regimes within our simulated forward search experiment. In the case of independent Gaussian priors over the regression coefficients, and in this specific case, it seems the asymptotic correlation may be closer to $-1$, as using a bias correction with magnitude $2$ fits the data best across all $n$ and $\rho$ regimes. Using smaller values (including the compromise value $1.5$ we eventually chose) are liable to under-estimate the bias in the case $n = p$. The difference between magnitudes lessens as $n$ grows, and we could already be comfortable using any of them once $n > 4p$ (again, in this lone data example). When the R2D2 prior is used the results are similar. The main difference we observe is that the difference between the three magnitudes is much smaller, regardless of the $n$ regime. The use of sparsity-inducing priors remedies selection-induced bias, and we do not have to be as concerned as in the high-risk regimes of independent Gaussian priors.

It is not clear how the magnitude in our bias estimate can be constructed generally in a principled manner. Equivalently, it is not clear how to estimate the asymptotic correlation between LOO-CV estimates and Bayes generalisation loss for a given data task, and we leave this as a promising area of future work.
\section{Sparsity-inducing priors}\label{sec:r2d2}
Fitting high-dimensional models to sparse data is generally challenging. When using joint shrinkage priors, practitioners explicit a prior belief that applies to both the parameter space and the predictive space. One such prior is the R2D2 prior \citep{zhang_bayesian_2022,aguilar_intuitive_2023,yanchenko_space_2023,yanchenko_r2d2_2023}, which allows statisticians to express prior beliefs about the model's $R^2$, and the number of predictors required to achieve it. By doing so, the risk of over-fitting the data is reduced compared to using simpler independent Gaussians for regression coefficients, which implicitly encode a (surprisingly) high \textit{a priori} $R^2$. The R2D2 model for a linear regression is:
\begin{align}
    y_i &\sim \normal(\beta_0 + \sum_{k=1}^{p}x_{k,i}\beta_k,\sigma^2) \\
  \beta_0 &\sim \studentt_3(0, 2.5) \\
  \sigma &\sim \studentt^{+}_3(0, 2.5) \\
  \beta_k &\sim \normal(0, \sigma^2 \tau^2 \phi_k) \\
  \tau^2 &= \frac{R^2}{1-R^2}\\
  R^2 &\sim \betadist(\mu_{R^2}, \varphi_{R^2}) \\
  \phi &\sim \Dirichlet(\xi,\dotsc,\xi).
\end{align}
We employ a mean and a pseudo-precision parameterisation of the beta distribution as opposed to shape parameters $a,b > 0$. The Dirichlet concentration parameter $\xi$, which is constant across all predictors, controls the sparsity in regression coefficients. Larger values of $\xi$ result in more uniformly distributed regression coefficients (in terms of magnitude), usually with more predictors explaining the response's variance. 

Alongside the R2D2 prior, \citet{piironen_sparsity_2017} propose the regularised horseshoe (RHS) prior. Here, we define the model as 
\begin{align}
  \beta_k &\sim \normal(0, \tau^2\tilde{\lambda}_k^2) \\
  \tilde{\lambda}_k^2 &= \frac{c^2\lambda_k^2}{c^2 + \tau^2\lambda_k^2} \\
  \lambda_k^2 &\sim \Cauchy^{+}(0, 1).
\end{align}
Across all binary classification case studies in this paper, we set the degrees of freedom of Student-$t$ prior of the global and local shrinkage parameters to be one, and the degrees of freedom of the slab to be $100$ so as to approach the horseshoe prior of \citet{carvalho_handling_2009}. Rather than define our prior over $\tau$ and $c$, we instead prefer to consider the number of effective non-zero parameters, denoted $m_\mathrm{eff}$, or rather the proportion of effectively non-zero parameters, $m_\mathrm{eff}/p$. Table~\ref{tab:r2d2-hyperparams} presents the hyper-parameters (for both the R2D2 and RHS priors) used in the real world case studies.\footnote{An interactive visualisation of the R2D2 prior is freely available at \url{https://solviro.shinyapps.io/R2D2_shiny/}.}
\begin{table}[!t]
    \centering
    \begin{tabular}{llcccccc}
        \toprule
        Data & Task & $n$ & $p$ & Prior & Hyper-parameters \\ \midrule
        Crime & Regression & $1992$ & $102$ & R2D2 & $\xi = 1;\, \mu_{R^2} = 0.2;\, \varphi_{R^2} = 5$ \\
        Simulated & 
        Regression & varies & $100$ & R2D2 & $\xi = 0.5;\, \mu_{R^2} = 0.5;\, \varphi_{R^2} = 4$ \\
        \midrule
        Heart & Classification & $224$ & $44$ & RHS & $m_\mathrm{eff}/p = 0.3$\\
        Ionosphere & Classification & $351$ & $33$ & RHS & $m_\mathrm{eff}/p = 0.6$\\
        Sonar & Classification & $208$ & $60$ & RHS & $m_\mathrm{eff}/p = 0.3$\\
        \bottomrule 
    \end{tabular}
    \caption{Specification of datasets, priors, and hyper-parameters used to produce the reference models throughout the paper.}
    \label{tab:r2d2-hyperparams}
\end{table}
\section{Alternative stopping rules}\label{sec:alternative-heuristics}
The previously presented stopping rules which involve choosing the ``bulge'' model, its associated $2\sigma$ rule, and projection predictive inference all suffer from the same flaw: they require running the forward search up to (at least) the bulge model. A simple alternative would be to run the forward search until none of the candidates improves predictive performance as judged by the two or three standard errors of their elpd difference estimates. Formally, we stop the search if there is \textit{no} candidate model, $\Mk$, such that
\begin{equation}
    \Delta\elpdHat{\Mk, \model_\text{previous}}{y} - 2\times\mathrm{se}\left(\Delta\elpdHat{\Mk, \model_\text{previous}}{y}\right) \geq 0,\label{eq:2-delta-sigma-rule}
\end{equation}
where $\model_\text{previous}$ denotes the model chosen in the previous step. Or using a magnitude of $3$ rather than $2$ in Equation~\ref{eq:2-delta-sigma-rule} to be even more conservative. Let us call these two rules $2\sigma_\Delta$ and $3\sigma_\Delta$ respectively. 

\begin{figure}[!t]
    \centering
    \includegraphics{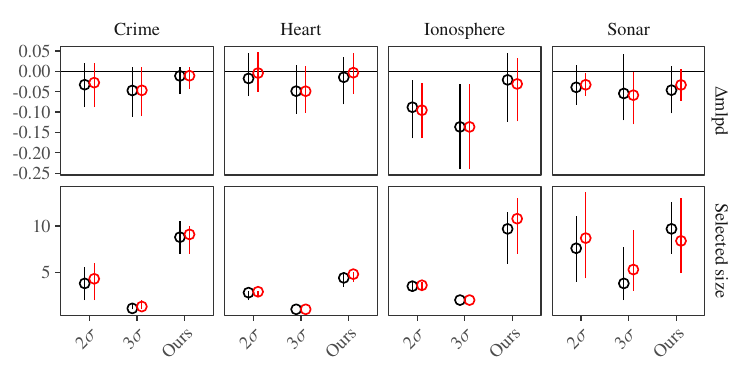}
    \caption{Real world datasets. The selected model sizes in three real world datasets across different heuristics when using standard independent Gaussian priors (in black) and the R2D2 prior (in \textcolor{red}{red}). Our correction identifies larger model sizes that the $2\sigma_\Delta$ and $3\sigma_\Delta$, but with higher mlpd point estimates. The incremental stopping rules are liable to select too-small models, resulting in sub-optimal out-of-sample performance and reduced intuition.}
    \label{fig:alternative-rules}
\end{figure}

In Figure~\ref{fig:alternative-rules} we show the model sizes they select, along with the test mlpd of those models selected. We find that these rules identify much smaller model sizes than our proposal, although do so at the cost of predictive performance. Our proposal finds models with higher mlpd point estimates across all datasets, and with sizes only slightly larger. Further, we find that the $2\sigma_\Delta$ and $3\sigma_\Delta$ rules can be volatile, and less stable than our proposal.

It is known that there types of heuristics (the $2\sigma_\Delta$ and $3\sigma_\Delta$ rules) struggle under collinearity of predictors, and in the presence of weakly (but truly) relevant predictors. This is seen in these real data cases, where the heuristics choose model sizes that, while being computationally beneficial, may be overly-sparse. For these reasons, we do not consider them to sufficiently robust (in a qualitative sense) for use in real data experiments where the structure of the predictors is unknown.
\end{appendices}
\end{document}